\documentclass[fleqn,usenatbib]{mnras}

\usepackage{newtxtext,newtxmath}

\usepackage[T1]{fontenc}

\DeclareRobustCommand{\VAN}[3]{#2}
\let\VANthebibliography\thebibliography
\def\thebibliography{\DeclareRobustCommand{\VAN}[3]{##3}\VANthebibliography}

\usepackage{graphicx}	
\usepackage{amsmath}	
\usepackage{CJK}

\newcommand   {\EBV}     {E(B\,{-}\,V)}
\newcommand   {\Teff}    {T_{\rm eff}}

\newcommand   {\feh}     {\rm [Fe/H]}
\newcommand   {\kms}     {\rm km\,s^{-1}}

\defcitealias{Planck2016dust}{Planck}

\title[DIBs: vertical distribution]{The Pristine Inner Galaxy Survey (PIGS) VI: Different vertical 
distributions between two DIBs at 442.8 nm and 862.1 nm}

\author[H. Zhao et al.]{
He Zhao (赵赫),$^{1,2}$\thanks{E-mail: he.zhao@oca.eu}
Mathias Schultheis,$^{2}$\thanks{E-mail: mathias.schultheis@oca.eu}
Anke Arentsen,$^{3,4}$
Georges Kordopatis,$^{2}$
Morgan Fouesneau,$^{5}$
\newauthor
Else Starkenburg,$^{6}$ 
Federico Sestito,$^{7}$
Vanessa Hill,$^{2}$
Nicolas F. Martin,$^{4,5}$
S\'{e}bastien Fabbro,$^{8}$
A.B.A. Queiroz$^{9,10}$
\\
$^{1}$Purple Mountain Observatory, Chinese Academy of Sciences, Nanjing 210023, PR China \\
$^{2}$Universit\'e C\^ote d'Azur, Observatoire de la C\^ote d'Azur, CNRS, Laboratoire Lagrange, Observatoire Bd, CS 34229, 06304 Nice cedex 4, France \\
$^{3}$Institute of Astronomy, University of Cambridge, Madingley Road, Cambridge CB3 0HA, UK \\
$^{4}$Universit\'e de Strasbourg, CNRS, Observatoire astronomique de Strasbourg, UMR 7550, F-67000 Strasbourg, France \\
$^{5}$Max-Planck-Institut f\"ur  Astronomie, K\"onigstuhl 17, D-69117 Heidelberg, Germany \\
$^{6}$Kapteyn Astronomical Institute, University of Groningen, Landleven 12, NL-9747 AD Groningen, the Netherlands \\
$^{7}$Department of Physics and Astronomy, University of Victoria, Victoria, BC V8W 3P2, Canada \\
$^{8}$National Research Council of Canada, Herzberg Astronomy \& Astrophysics Program, 5071 West Saanich Road, Victoria, BC, V9E 2E7, Canada \\
$^{9}$Leibniz-Institut f\"ur Astrophysik Potsdam (AIP), An der Sternwarte 16, 14482 Potsdam, Germany \\
$^{10}$Institut f\"{u}r Physik und Astronomie, Universit\"{a}t Potsdam, Haus 28 Karl-Liebknecht-Str. 24/25, D-14476 Golm, Germany
}
\date{Accepted 2022 November 25. Received 2022 November 25; in original form 2022 September 28}

\pubyear{2022}

\begin{document}
\begin{CJK*}{UTF8}{gbsn}

\label{firstpage}
\pagerange{\pageref{firstpage}--\pageref{lastpage}}
\maketitle

\begin{abstract}
Although diffuse interstellar bands (DIBs) were discovered over 100 years ago, for most of them, their origins are still unknown.
Investigation on the correlations between different DIBs is an important way to study the behavior and distributions of their carriers. 
Based on stacking thousands of spectra from the Pristine Inner Galaxy Survey, we study the correlations between two DIBs at 442.8\,nm 
($\lambda$442.8) and 862.1\,nm ($\lambda$862.1), as well as the dust grains, in a range of latitude spanning ${\sim}22^{\circ}$
($4\degr\,{<}\,|b|\,{<}\,15\degr$) toward the Galactic center ($|\ell|\,{<}\,11^{\circ}$). Tight linear intensity correlations 
can be found between $\lambda$442.8, $\lambda$862.1, and dust grains for $|b|\,{<}\,12^{\circ}$ or $\EBV\,{>}\,0.3$\,mag. 
For $|b|\,{>}\,12^{\circ}$, $\lambda$442.8 and $\lambda$862.1 present larger relative strength with respect to the dust grains.
A systematic variation of the relative strength between $\lambda$442.8 and $\lambda$862.1 with $|b|$ and $\EBV$ concludes that the 
two DIBs do not share a common carrier. Furthermore, the carrier of $\lambda$862.1 is more abundant at high latitudes than that of 
$\lambda$442.8. This work can be treated as an example showing the significance and potentials to the DIB research covering a 
large latitude range.
\end{abstract}






\section{Introduction} 

After over 100 years since the discovery of the diffuse interstellar bands (DIBs) in 1919 \citep{Heger1922}, over 600 DIBs have
been confirmed between 0.4 and 2.4\,{\micron} \citep{Cox2014,Galazutdinov2017b,Fan2019,Hamano2022,Ebenbichler2022}. As a set of weak 
and broad absorption features, today DIBs are thought to be produced by carbon-bearing molecules, like carbon or hydrocarbon chains 
\citep[e.g.,][]{Maier2011a,ZM2014iaus}, polycyclic aromatic hydrocarbons \citep[PAHs; e.g.,][]{Salama1996,Shen2018,Omont2019}, and 
fullerenes and their derivatives \citep[e.g.,][]{Fulara1993a,Cami2014iaus,Omont2016}. However, due to the difficulties in the experimental 
research on complex molecules \citep{Hardy2017,Kofman2017} and in the comparison between the experimental measurements and astronomical 
observations, buckminsterfullerene $(C_{60}^{+})$ is the first and only identified DIB carrier for five near-infrared DIBs so far 
\citep[e.g.,][]{FE1994,Campbell2015,Walker2016,Linnartz2020}, although some debates about the wavelength match and the relative 
strength of these bands still exist \citep{Galazutdinov2017a,Galazutdinov2021}.

Besides the comparison between astronomical observations and experimental results, investigating the correlations between different
DIBs is also one of the most important ways to study the relations between their carriers and even to find the common carrier for a set 
of DIBs \citep[e.g.,][]{Friedman2011,Ensor2017,Elyajouri2017b,Elyajouri2018}. The tightest correlation was found between two DIBs at
619.6\,nm and 661.4\,nm (in this work, we cite DIBs with their central wavelengths in nanometer) with very high Pearson coefficient
($r_p\,{>}\,0.98$; e.g., \citealt{McCall2010}; \citealt{Friedman2011}; \citealt{KZ2013}; \citealt{Bondar2020}). But the variation of
their strength ratio has also been reported by \citet{Krelowski2016} and \citet{Fan2017}, verifying that only a tight intensity
correlation is not enough to conclude a common origin for different DIBs. The behavior of the relative strength between different
DIBs as a function of $f_{\rm H_2}\,{\equiv}\,2N({\rm H_2})/[N(\ion{H}{i}) + 2{N}({\rm H_2})]$ was used by \citet{Fan2017} to study 
the relative positions of the DIB carriers. \citet{Lan2015} also investigated the correlation between DIB strength and $N(\ion{H}{i})$
and $N({\rm H_2})$ for 20 DIBs at high latitudes. Nevertheless, it is hard to measure $N(\ion{H}{i})$ and $N({\rm H_2})$ in ultraviolet
spectra or decipher the positions of \ion{H}{i} and $\rm H_2$ from the radio observations. Another way is to explore the spatial
distributions of different DIBs, which requires the probing of interstellar medium (ISM) above and below the Galactic plane over a large range of 
latitudes if we would like to get more information and conclusions from the correlation study on different DIBs. \citet{MW1993} 
studied the variation of the relative strength between DIBs $\lambda$442.8 , $\lambda$578.0, and $\lambda$579.7 as a function of Galactic latitudes 
based on a sample of 65 stars. They found the carrier abundance of $\lambda$442.8 to be highest at low latitudes,
which agrees with our results (see Sect. \ref{subsect:distribution2}).


The DIB research benefits from the arrival of large spectroscopic surveys allowing to perform large  statistically studies. The 
three-dimensional (3D) distributions of the DIB carriers have been mapped by \citet{Kos2014} and \citet{Schultheis2022} for DIB\,$\lambda$862.1
(we take here its central wavelength as 862.086\,nm measured in \citealt{Schultheis2022}), and \citet{Zasowski2015c} for
DIB\,$\lambda$1527.3, based on the data from the Radial Velocity Experiment \citep[RAVE;][]{Steinmetz2006}, {\it 
Gaia}--DR3 \citep{Vallenari2022}, and the Apache Point Observatory Galactic Evolution Experiment \citep[APOGEE;][]{Majewski2017}, respectively.
\citet{Zasowski2015c}, \citet{hz2021b}, and \citet{Schultheis2022} made preliminary studies on the kinematics of the DIB carriers 
with the APOGEE, {\it Gaia}--ESO \citep{Gilmore2012}, and {\it Gaia} data sets. Nevertheless, the individual spectra in large 
spectroscopic surveys usually have less integral time than specifically designed DIB observations, resulting in a lower signal-to-noise 
ratio ($\rm S/N$). Taking this into account, stacking spectra in an arbitrary spatial volume is a practical and useful method to achieve 
better $\rm S/N$ and to precisely measure DIB features \citep[e.g.,][]{Kos2013,Lan2015,Baron2015a,Baron2015b}. 

Based on the survey data, some studies devoted to the investigation on the intensity correlations between different DIBs.
\citet{Elyajouri2017b} made use of $\sim$300 spectra of early-type stars in APOGEE to explore the correlations between the strong 
DIB at 1.5273\,{\micron} and the three weak DIBs at 1.5627, 1.5653, and 1.5673\,{\micron}. A comparison between the DIB at 1.5273\,{\micron}
and some optical DIBs was done as well. Based on 250 stacked spectra at high latitudes, \citet{Baron2015b} successfully clustered
26 weak DIBs into six groups and four of them were tightly associated with $\rm C_2$ or CN. A data-driven analysis was also done 
by \citet{Fan2022} for 54 strong DIBs measured in 25 high-quality spectra of early-type stars. And they suggested a continuous
change of properties of the DIB carriers between different groups. The results of \citet{Puspitarini2015} showed a similar variation 
of the strength with the distance of background stars for DIBs $\lambda$661.4 and $\lambda$862.1 in a field centered at 
$(\ell,b)\,{=}\,(212.9^{\circ},{-}2.0^{\circ})$. But a direct comparison between $\lambda$661.4 and $\lambda$862.1 was not made. 

In this work, we take advantage of the data from the metal-poor Pristine Inner Galaxy Survey \citep[PIGS;][]{Arentsen2020b} which
contain a large number of spectra (13\,235) and two strong DIBs ($\lambda$442.8 and $\lambda$862.1) in its blue-band and red-band 
spectra, respectively. Metal-poor stars have the advantage that the DIBs are less or not at all affected by stellar lines. We 
measure the two DIBs in stacked spectra and investigate their relative vertical distributions. In Sect. \ref{sect:pigs}, we briefly 
introduce the PIGS survey. The stacking of spectra and the DIB measurements are described in Sect. \ref{sect:fit-dib}. The results 
of the intensity correlations and vertical distributions of the two DIBs and dust grains are presented in Sect. \ref{sect:result} 
and discussed in Sect. \ref{sect:discuss}. The main conclusions are summarized in Sect. \ref{sect:conclusion}.

\begin{figure}
    \centering
    \includegraphics[width=8cm]{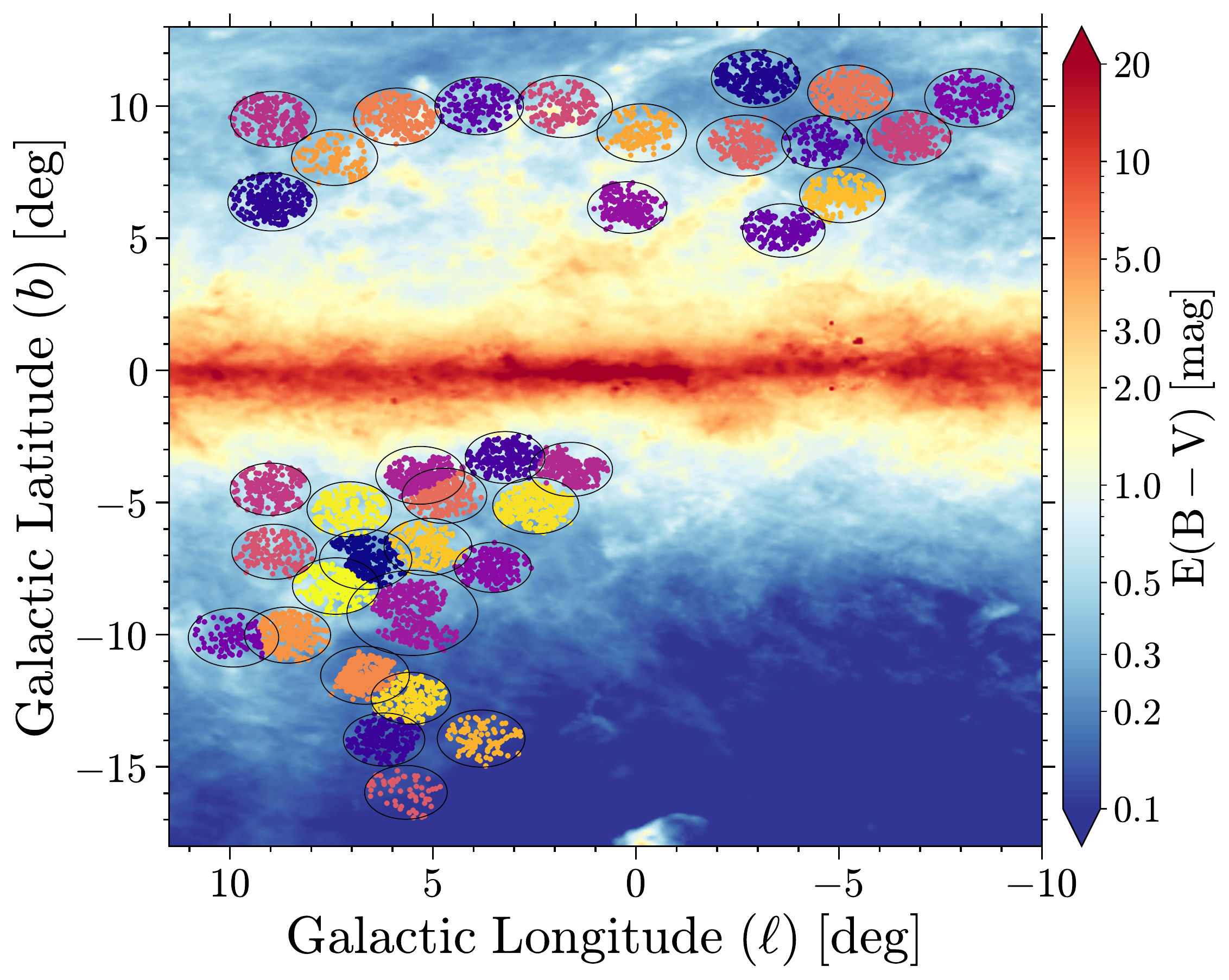}
    \caption{Spatial distribution ($\ell,b$) of 6980 PIGS targets, overplotted on the dust reddening map of \citet{Planck2016dust}.
    Colored dots represent the targets assigned into different fields (black circles), as a result of the k-means clustering with 
    $N=36$ (see Sect. \ref{sect:fit-dib} for details).}
    \label{fig:field}
\end{figure}

\section{Pristine Inner Galaxy Survey (PIGS)} \label{sect:pigs}

The Pristine Inner Galaxy Survey \citep[PIGS;][]{Arentsen2020b,Arentsen2020a,Arentsen2021} is an extension of the Pristine survey, 
which uses the metallicity-sensitive narrow-band $CaHK$ filter on the Canada-France-Hawaii-Telescope (CFHT) to search for and study 
the most metal-poor stars \citep{Starkenburg2017}. PIGS aims at obtaining spectra for the metal-poor stars in the Galactic bulge 
and studying their kinematics \citep{Arentsen2020a}, as well as the chemical and dynamical evolution of the inner Galaxy \citep{Arentsen2020b,
Arentsen2021,Sestito2022}. The PIGS targets were selected with a magnitude limit of $13.5\,{<}\,G\,{<}\,16.5$\,mag for {\it Gaia} \citep{Gaia2018} 
or $14.0\,{<}\,g\,{<}\,17.0$\,mag for Pan--STARRS1 \citep{Chambers2016}, and an reddening limit of $\EBV\,{\lesssim}\,0.7$\,mag 
from \citet{Green2018}. Most of these targets (88\%) have $\feh\,{<}\,{-}1.0$\,dex, with a peak around --1.5\,dex and a tail down 
to --3.0\,dex \citep{Arentsen2020b}. The targets were observed with AAOmega+2dF on the AAT, obtaining simultaneous blue-band
(370--550\,nm, $R\,{\sim}\,1300$) and red-band (840--880\,nm, $R\,{\sim}\,11\,000$) spectra. 

The spectra were analyzed with the FERRE\footnote{FERRE \citep{Allende2006} is available from \url{http://github.com/callendeprieto/ferre}} 
code, which simultaneously derived effective temperatures, surface gravities, metallicities, and carbon abundances. For details on 
the analysis, see \citet{Arentsen2020b}. In the original analysis, both the observed and model spectra were normalized using a 
running mean. For this work, we perform a re-normalization of the original observed and best-fitting synthetic spectra using the 
\textit{fit-continuum} task in the Python {\it specutils} package. A third and fifth order Chebyshev polynomial was used for 
red-band and blue-band spectra, respectively. 

There are 13\,235 PIGS spectra observed between 2017 and 2020, of which we make use of 6980 of them, distributed into 36 fields 
(see Fig. \ref{fig:field}), with $\rm S/N\,{>}\,50$ measured between 840--880\,{nm} and $\Teff\,{<}\,7000$\,K, which assures the 
quality of the observed and synthetic spectra. For this subsample, $\rm S/N$ is mostly below 150 per pixel for red-band (computed 
between 840--880\,{nm}) with a mean of 77, and below 50 per pixel for blue-band (computed between 400--410\,{nm}) with a mean of 30. 
Thus, in this work, we only fit and measure two strong DIBs, $\lambda$442.8 and $\lambda$862.1, in stacked blue-band and red-band 
spectra, respectively, due to the low $\rm S/N$ of individual PIGS spectra. We applied a simple k-means algorithm \citep{Lloyd1982} 
to cluster targets into different fields to avoid the possible overlap of observed PIGS fields, especially in the southern footprint,
and to have a cleaner selection of target stars in the same $(\ell,\,b)$ range, because an overlap of fields would smooth the variation of
dust reddening and DIB strength with $(\ell,\,b)$. In some cases it is also helpful for fields with worse quality spectra, to have a 
larger number of stars, such as the field at $(\ell_0,\,b_0)=(9.91^{\circ},\,{-}10.11^{\circ})$. The clustering was completed by the Python {\it 
scikit-learn} package \citep{scikit-learn} with $N=36$, and the result is shown in Fig. \ref{fig:field}. In the following analysis, 
the PIGS ``fields'' refer to the assigned clustered regions, which follow but are not exactly the same as their observational footprints. 
Discrete footprints are well clustered (e.g., the targets at $5^{\circ}\,{<}\,b\,{<}\,12^{\circ}$), while in crowded regions, such 
as the targets around $(\ell,\,b)=(5^{\circ},\,{-}8^{\circ})$, the clustered fields may be different from the observational 
ones. The central coordinates $(\ell_0,b_0)$, radius, and target number of each field are listed in Table \ref{tab:stack-fit4430}.

\begin{table*}
	\begin{center}
	\normalsize
	\caption{Field information and fitting results of DIB\,$\lambda$442.8 in the blue-band stacked ISM spectra. \label{tab:stack-fit4430}}
		\begin{tabular}{l r r c r c r c c c }
		\hline\hline 
Field & $\ell_0$   & $b_0$      & radius     & $N^a$ & $\EBV^b$ & $\rm S/N^c$  & $\lambda_C^d\pm{\rm err}$ & $\rm FWHM^e\pm{\rm err}^e$ & $\rm EW_{fit}^f \pm err$ \\
Nr    & ($^\circ$) & ($^\circ$) & ($^\circ$) &       & (mag)    &              & (nm)                    & (nm)                       & ({\AA}) \\ [0.5ex]
\hline
 1 &   6.66	&  --7.15 &	1.14 & 259 & $0.37\pm0.05$ & 145.5 & $442.76^{+0.08}_{-0.08}$ & $1.99^{+0.18}_{-0.16}$ & $1.10\pm0.08$ \\ [0.5ex]
 2 & --2.94 &   11.04 &	1.09 & 257 & $0.31\pm0.05$ & 104.7 & $442.71^{+0.10}_{-0.09}$ & $2.01^{+0.18}_{-0.18}$ & $0.93\pm0.09$ \\ [0.5ex]
 3 &   8.96	&    6.37 &	1.10 & 236 & $0.57\pm0.09$ & 119.0 & $442.79^{+0.07}_{-0.08}$ & $1.83^{+0.16}_{-0.16}$ & $1.53\pm0.09$ \\ [0.5ex]
 4 &   6.20	& --13.97 &	1.00 & 196 & $0.14\pm0.02$ & 119.7 & $442.54^{+0.10}_{-0.10}$ & $2.07^{+0.18}_{-0.18}$ & $0.73\pm0.10$ \\ [0.5ex]
 5 &   3.22	&  --3.30 &	0.98 & 224 & $0.66\pm0.10$ &  89.7 & $442.68^{+0.08}_{-0.08}$ & $2.18^{+0.18}_{-0.18}$ & $1.98\pm0.16$ \\ [0.5ex]
 6 & --4.59 &	 8.63 &	1.00 & 104 & $0.26\pm0.06$ & 101.7 & $442.53^{+0.10}_{-0.10}$ & $2.23^{+0.20}_{-0.18}$ & $0.99\pm0.11$ \\ [0.5ex]
 7 &   3.86	&   10.00 &	1.09 & 135 & $0.40\pm0.10$ &  93.5 & $442.77^{+0.09}_{-0.09}$ & $2.03^{+0.18}_{-0.16}$ & $1.13\pm0.10$ \\ [0.5ex]
 8 & --3.64 &    5.29 &	1.02 & 180 & $0.68\pm0.09$ & 112.6 & $442.73^{+0.08}_{-0.08}$ & $2.08^{+0.16}_{-0.16}$ & $1.84\pm0.10$ \\ [0.5ex]
 9 &   9.91	& --10.11 &	1.11 &  99 & $0.32\pm0.07$ & 141.4 & $442.71^{+0.09}_{-0.09}$ & $2.40^{+0.20}_{-0.18}$ & $1.13\pm0.09$ \\ [0.5ex]
10 & --8.22 &   10.31 &	1.11 & 185 & $0.29\pm0.04$ & 122.5 & $442.52^{+0.09}_{-0.09}$ & $2.00^{+0.18}_{-0.18}$ & $0.92\pm0.09$ \\ [0.5ex]
11 &   3.53	&  --7.45 &	0.95 & 203 & $0.33\pm0.03$ & 128.8 & $442.65^{+0.09}_{-0.08}$ & $2.12^{+0.18}_{-0.18}$ & $1.25\pm0.12$ \\ [0.5ex]
12 &   0.22	&    6.16 &	0.98 & 225 & $0.75\pm0.07$ & 105.7 & $442.81^{+0.07}_{-0.07}$ & $1.97^{+0.16}_{-0.16}$ & $2.11\pm0.12$ \\ [0.5ex]
13 &   5.51	&  --9.18 &	1.61 & 292 & $0.27\pm0.07$ & 133.7 & $442.74^{+0.08}_{-0.08}$ & $1.93^{+0.18}_{-0.18}$ & $1.23\pm0.09$ \\ [0.5ex]
14 &   5.32	&  --3.97 &	1.09 & 249 & $0.70\pm0.10$ & 117.4 & $442.75^{+0.07}_{-0.07}$ & $2.10^{+0.16}_{-0.16}$ & $2.01\pm0.10$ \\ [0.5ex]
15 &   8.93	&    9.50 &	1.06 & 182 & $0.39\pm0.04$ & 126.2 & $442.66^{+0.09}_{-0.09}$ & $2.05^{+0.18}_{-0.16}$ & $1.24\pm0.09$ \\ [0.5ex]
16 &   9.00	&  --4.50 &	0.99 & 183 & $0.61\pm0.11$ &  98.2 & $442.72^{+0.09}_{-0.09}$ & $2.26^{+0.18}_{-0.18}$ & $1.77\pm0.12$ \\ [0.5ex]
17 &   1.76	&    9.98 &	1.18 & 146 & $0.59\pm0.16$ & 122.3 & $442.82^{+0.08}_{-0.09}$ & $2.05^{+0.18}_{-0.18}$ & $1.20\pm0.10$ \\ [0.5ex]
18 &   8.91	&  --6.86 &	1.04 & 130 & $0.39\pm0.08$ & 133.4 & $442.68^{+0.09}_{-0.09}$ & $2.13^{+0.18}_{-0.16}$ & $1.35\pm0.10$ \\ [0.5ex]
19 & --5.27 &   10.50 &	1.05 & 212 & $0.25\pm0.03$ & 121.7 & $442.65^{+0.10}_{-0.10}$ & $1.86^{+0.18}_{-0.18}$ & $0.72\pm0.08$ \\ [0.5ex]
20 &   5.87	&    9.60 &	1.08 & 238 & $0.53\pm0.18$ & 127.7 & $442.68^{+0.08}_{-0.08}$ & $2.14^{+0.18}_{-0.16}$ & $1.40\pm0.11$ \\ [0.5ex]
21 &   8.58	& --10.02 &	1.06 & 250 & $0.36\pm0.06$ & 130.4 & $442.72^{+0.09}_{-0.09}$ & $1.95^{+0.18}_{-0.18}$ & $0.95\pm0.06$ \\ [0.5ex]
22 &   7.42	&    8.06 &	1.06 & 104 & $0.46\pm0.10$ & 112.3 & $442.74^{+0.09}_{-0.09}$ & $1.94^{+0.16}_{-0.16}$ & $1.33\pm0.09$ \\ [0.5ex]
23 & --0.14 &    8.98 &	1.10 & 136 & $0.52\pm0.12$ & 120.6 & $442.71^{+0.08}_{-0.08}$ & $1.97^{+0.18}_{-0.16}$ & $1.49\pm0.10$ \\ [0.5ex]
24 & --5.09 &    6.64 &	1.06 & 177 & $0.42\pm0.12$ & 121.9 & $442.67^{+0.08}_{-0.09}$ & $2.18^{+0.18}_{-0.18}$ & $1.31\pm0.10$ \\ [0.5ex]
25 &   5.12	&  --6.68 &	1.08 & 151 & $0.35\pm0.05$ & 128.1 & $442.66^{+0.09}_{-0.09}$ & $1.97^{+0.18}_{-0.16}$ & $1.22\pm0.09$ \\ [0.5ex]
26 &   5.54	& --12.41 &	0.98 & 185 & $0.15\pm0.02$ & 135.8 & $442.62^{+0.09}_{-0.10}$ & $1.98^{+0.18}_{-0.18}$ & $0.74\pm0.07$ \\ [0.5ex]
27 &   2.47	&  --5.12 &	1.06 & 279 & $0.40\pm0.08$ & 140.5 & $442.78^{+0.08}_{-0.08}$ & $2.24^{+0.18}_{-0.16}$ & $1.54\pm0.09$ \\ [0.5ex]
28 &   7.06	&  --5.25 &	1.04 & 179 & $0.54\pm0.06$ & 102.4 & $442.68^{+0.08}_{-0.08}$ & $2.25^{+0.18}_{-0.18}$ & $1.71\pm0.13$ \\ [0.5ex]
29 &   7.39	&  --8.16 &	1.07 & 289 & $0.37\pm0.09$ & 139.7 & $442.80^{+0.08}_{-0.08}$ & $1.92^{+0.18}_{-0.16}$ & $1.14\pm0.07$ \\ [0.5ex]
			\hline
			\multicolumn{10}{l}{$^a$ The number of spectra used for stacking in each field.} \\
			\multicolumn{10}{l}{$^b$ Median $\EBV$ $\pm$ its standard deviation in each field derived from the \citetalias{Planck2016dust} map.} \\
            \multicolumn{10}{l}{$^c$ Signal-to-noise ratio of the stacked blue-band ISM spectra.} \\
            \multicolumn{10}{l}{$^d$ Measured central wavelength in the heliocentric frame.} \\
            \multicolumn{10}{l}{$^e$ Full width at half maximum of DIB\,$\lambda$442.8.} \\
            \multicolumn{10}{l}{$^f$ Fitted equivalent width of DIB\,$\lambda$442.8.} \\
		\end{tabular}
	\end{center}
\end{table*}

\begin{figure}
    \centering
    \includegraphics[width=7cm]{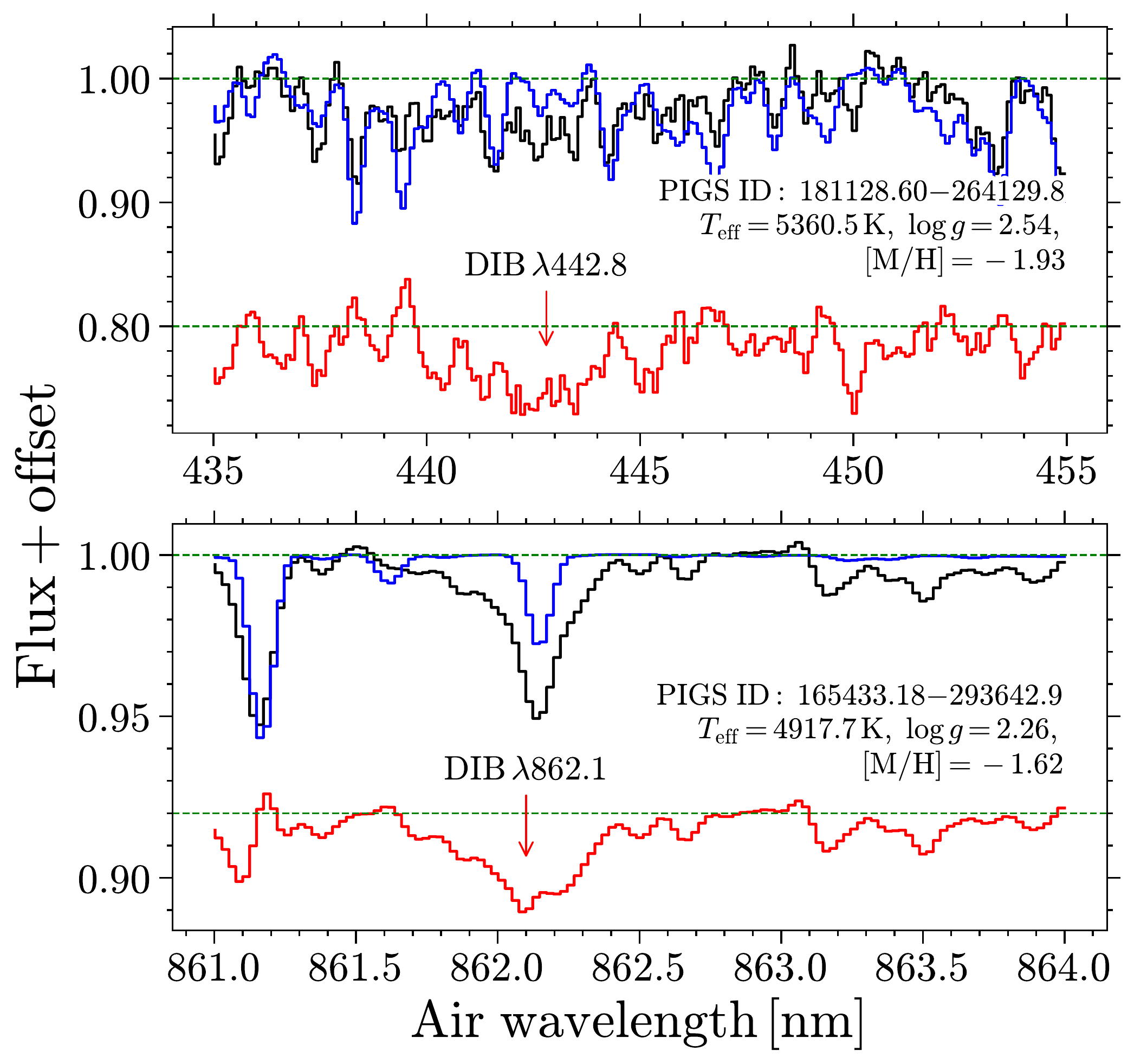}
    \caption{Two examples showing the DIB signals of $\lambda$442.8 ({\it upper panel}) and $\lambda$862.1 ({\it lower panel}) in their
    ISM spectra, respectively, derived by the observed blue-band and red-band PIGS spectra (black lines) and the corresponding synthetic spectra
    (blue lines). The DIB positions are marked. The PIGS ID and stellar atmospheric parameters of the two background stars are 
    indicated as well.}
    \label{fig:ISeg}
\end{figure}

\begin{figure}
    \centering
    \includegraphics[width=7cm]{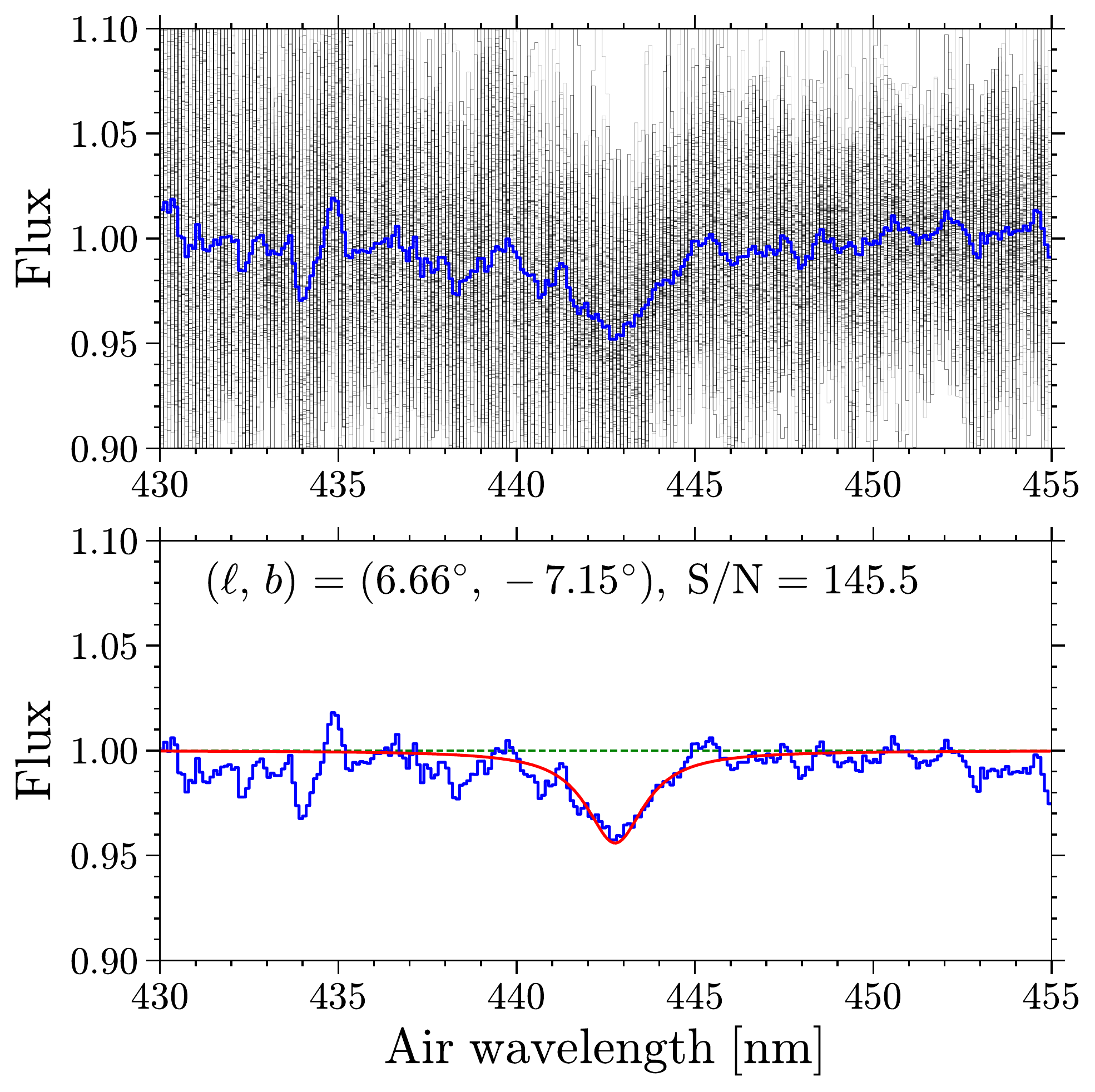}
    \caption{{\it Upper panel}: Example of stacking ISM spectra in the field $\left(\ell_0,b_0\right)=\left(6.66^{\circ},{-}7.15^{\circ}\right)$. 
    Black lines are individual ISM spectra, and the blue line is the stacked ISM spectrum. {\it Lower panel}: Fit of DIB\,$\lambda$442.8 
    in the stacked ISM spectrum after local renormalization (blue line). The red line shows the fitted Lorentzian profile. Field 
    position and $\rm S/N$ of the stacked ISM spectrum are indicated.}
    \label{fig:fit-4430}
\end{figure}

\begin{figure}
    \centering
    \includegraphics[width=7cm]{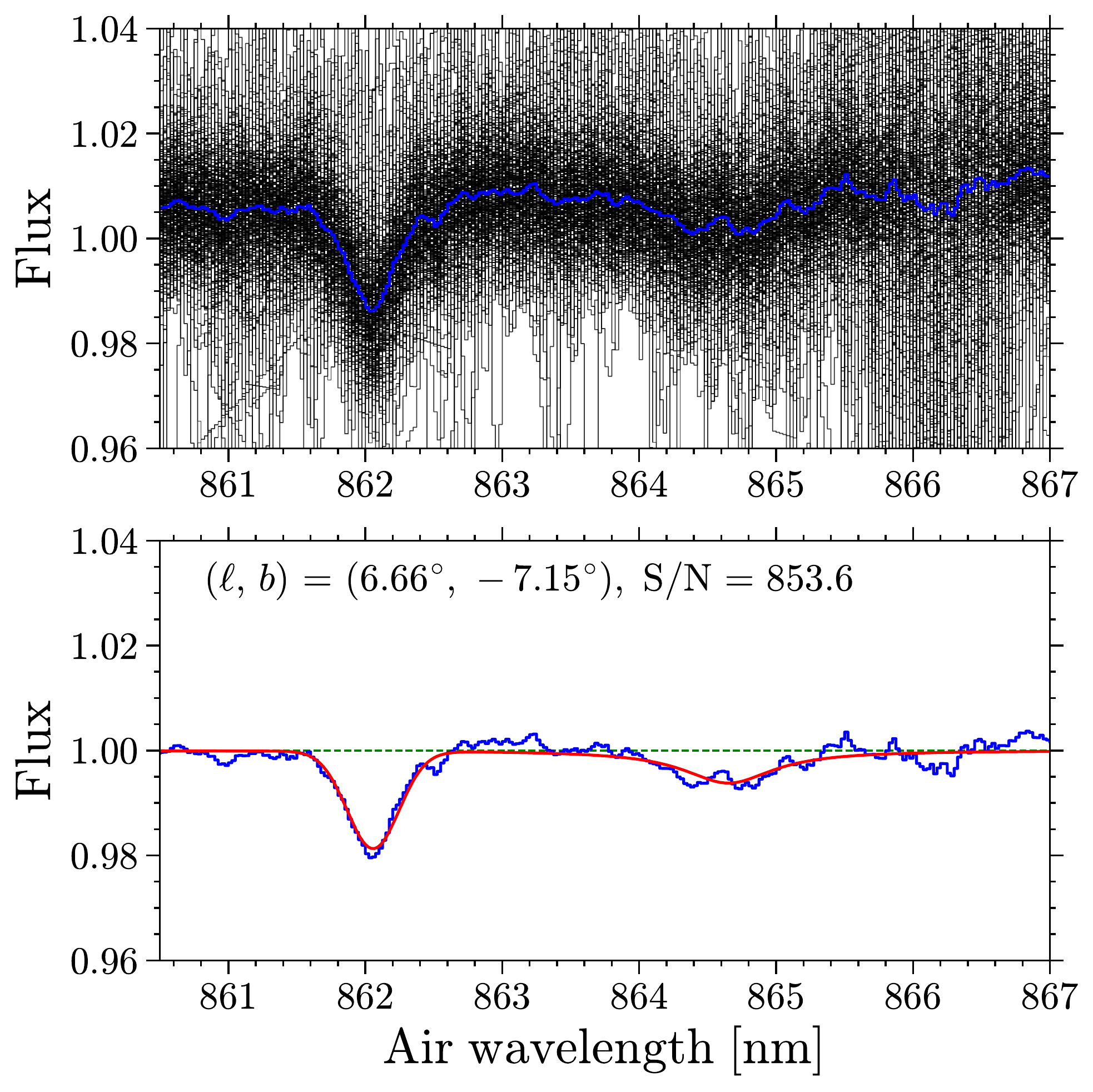}
    \caption{The same as Fig. \ref{fig:fit-4430}, but for DIB\,$\lambda$862.1.}
    \label{fig:fit-8620}
\end{figure}

\section{Fit and measure DIBs in stacked spectra} \label{sect:fit-dib}

Limited by the PIGS sample size and the low S/N of individual spectra, we choose to stack spectra in each field according 
to their Galactic coordinates $(\ell,b)$ without taking the stellar distance into account. Thus the DIB measured in the stacked 
spectra is a measure of the average column density of its carrier toward a given sightline. 

Before stacking spectra in each field, the stellar components in observed spectra are first subtracted by the synthetic spectra,
providing the ISM spectra for each target. Figure \ref{fig:ISeg} shows two ISM spectra derived from the blue-band and red-band observed 
spectra, respectively, subtracted by their synthetic spectra. The DIB signals are clear, although their profiles are contaminated by 
noise and the residuals of stellar lines, such as the \ion{Fe}{i} line close to the center of $\lambda$862.1. We emphasize 
that in the stacked ISM spectra these contamination are significantly alleviated (see Figs. \ref{fig:fit-4430} and \ref{fig:fit-8620}) 
due to the averaging of a substantial number of spectra and the large velocity dispersion of stars in each field (standard deviation 
${>}100\,\kms$). The second step is to shift the ISM spectra back to the heliocentric frame using the stellar radial velocities 
(${\rm RV_{star}}$, in $\kms$), that is $\lambda_{\rm pixel}^{\prime} = \lambda_{\rm pixel} + {\rm RV_{star}} \times \lambda_{\rm pixel}/c$, 
where $\lambda_{\rm pixel}^{\prime}$ and $\lambda_{\rm pixel}$ are the wavelength pixels in the heliocentric and stellar frames, 
respectively, and $c\,{=}\,3\,{\times}\,10^5\,\kms$ is the speed of light. Finally, stacking of individual ISM spectra in each field 
is done for the blue band and red band, respectively, by taking the median value of their flux which could reduce the influence 
of the outlier pixels and discrepancy between the individual observed and synthetic spectra. A $\rm S/N$ is calculated between 860.5 
and 861.5\,nm for red-band stacked ISM spectra by $\rm mean(flux)/std(flux)$. For blue-band stacked ISM spectra, fluxes are used in 
two windows, that is 430--433\,nm and 452--455\,nm. Examples for blue-band and red-band stacked ISM spectra can be found in the upper 
panels in Figs. \ref{fig:fit-4430} and \ref{fig:fit-8620}, respectively. The $\rm S/N$ of stacked spectra (listed in Tables 
\ref{tab:stack-fit4430} and \ref{tab:stack-fit8620}) are dramatically increased. 

Compared to the red band, the blue-band stacked ISM spectra are more noisy and have much more residual features with stellar 
origins. This could partly be due to the abundance variations which are common at low metallicity. Thus, fitting a continuum straight
through the ISM spectra could lead to an underestimation of the strength of $\lambda$442.8. Therefore, we apply the Gaussian process
regression \citep[GPR;][]{GB12,RW06} to fit the profile of $\lambda$442.8, the stellar residuals, and the random noise simultaneously. 
This method has been successfully applied to the spectra of early-type stars \citep{Kos2017,hz2021a}. Specifically, the blue-band 
stacked ISM spectra are first locally renormalized with the spectral window of 430--455\,nm by an iterated method using a second-order 
polynomial (see Sect. 2.2 in \citealt{hz2021a} for details). Then the ISM spectra are initially fitted by a Lorentzian function for
the profile of $\lambda$442.8, as suggested by \citet{Snow2002b}, and a constant continuum. Finally, GPR is applied with a Lorentzian
function as its mean function and a Mat\'{e}rn 3/2 kernel to model the correlated noise, including both the stellar residuals 
and random noise. Only one kernel is used here because DIB\,$\lambda$442.8 is the broadest feature in the ISM spectra. The priors
of fitting parameters mainly follow those in \citet{Kos2017} and \citet{hz2021a}, that is Gaussian priors centered at the initial 
fitting results with a width of 0.15\,nm for the DIB central wavelength and the Lorentzian width and a flat prior for the scale 
length of the Mat\'{e}rn 3/2 kernel. Details about GPR and the priors can be found in \citet{Kos2017} and \citet{hz2021a}.

The stellar residuals are much less in the red-band stacked ISM spectra. Among them, the strongest one is from the 
\ion{Ca}{ii} line around 866.5\,nm. But it is far away from the DIB\,$\lambda$862.1. Using stacked {\it Gaia}--RVS spectra 
\citep{Seabroke2022}, \citet{hz2022a} confirmed that the weak DIB around 864.8\,nm is very broad, with a Full Width at Half Maximum 
(FWHM) of $\sim$1.6\,nm, and its profile could affect the placement of the continuum (see Fig. 1 in \citealt{hz2022a}). Therefore, 
we fit each red-band stacked ISM spectrum between 860.5 and 867\,nm with a Gaussian function for the profile of $\lambda$862.1, a 
Lorentzian function for the profile of $\lambda$864.8, and a linear continuum. Nevertheless, the DIB\,$\lambda$864.8 profile cannot 
be well described due to the lower quality of the individual ISM spectra at longer wavelength (see upper panel in Fig. \ref{fig:fit-8620}). 
We emphasize that because DIBs are absorption features, the usage of the whole spectra between 860.5 and 867\,nm can help us to get 
a better placement of the continuum, although the $\lambda$864.8 profile cannot be well fitted. 

The parameter optimization, for both red band and blue band, is implemented by a Markov Chain Monte Carlo (MCMC) procedure 
\citep{Foreman-Mackey13}. The 50\% values in the posterior distribution generated by MCMC are treated as the best estimates, with 
lower and upper errors derived from the differences of 16\% and 84\% to 50\% values, respectively. Fit examples of $\lambda$442.8 
and $\lambda$862.1 are shown in Figs. \ref{fig:fit-4430} and \ref{fig:fit-8620}, respectively. 

\begin{table}
	\begin{center}
	\small 
	\caption{Fitting results of DIB\,$\lambda$862.1 in the red-band stacked ISM spectra. The field numbers are the same as Table 
	\ref{tab:stack-fit4430}. \label{tab:stack-fit8620}}
		\begin{tabular}{l c c c c c}
		\hline\hline 
			Field & $\rm S/N^a$  & $\lambda_C^b\pm{\rm err}$ & $\rm FWHM^c\pm err$ & $\rm EW_{fit}^d \pm err$ & $\rm EW_{int}^e$ \\
			Nr    &              & (nm)                      & (nm)                & ({\AA})                  & ({\AA})          \\ [0.5ex]
			\hline
			 1 & 853.6 & $862.06^{+0.02}_{-0.02}$ & $0.45^{+0.06}_{-0.05}$ & $0.088\pm0.004$ & 0.091 \\ [0.5ex] 
			 2 & 555.0 & $862.07^{+0.02}_{-0.02}$ & $0.46^{+0.06}_{-0.05}$ & $0.091\pm0.003$ & 0.093 \\ [0.5ex]
			 3 & 756.1 & $862.06^{+0.02}_{-0.02}$ & $0.48^{+0.05}_{-0.05}$ & $0.157\pm0.006$ & 0.159 \\ [0.5ex]
			 4 & 996.9 & $862.06^{+0.03}_{-0.02}$ & $0.56^{+0.08}_{-0.07}$ & $0.105\pm0.006$ & 0.107 \\ [0.5ex]
			 5 & 529.5 & $862.06^{+0.02}_{-0.02}$ & $0.47^{+0.05}_{-0.04}$ & $0.189\pm0.004$ & 0.190 \\ [0.5ex]
			 6 & 519.3 & $862.04^{+0.03}_{-0.03}$ & $0.46^{+0.08}_{-0.07}$ & $0.085\pm0.005$ & 0.091 \\ [0.5ex]
			 7 & 930.8 & $862.06^{+0.02}_{-0.02}$ & $0.47^{+0.05}_{-0.05}$ & $0.105\pm0.004$ & 0.111 \\ [0.5ex]
			 8 & 595.4 & $862.04^{+0.02}_{-0.02}$ & $0.46^{+0.05}_{-0.05}$ & $0.187\pm0.005$ & 0.191 \\ [0.5ex]
			 9 & 447.1 & $862.05^{+0.03}_{-0.03}$ & $0.42^{+0.07}_{-0.06}$ & $0.071\pm0.004$ & 0.072 \\ [0.5ex]
			10 & 583.6 & $861.96^{+0.02}_{-0.02}$ & $0.48^{+0.06}_{-0.05}$ & $0.094\pm0.004$ & 0.093 \\ [0.5ex]
			11 & 722.8 & $862.08^{+0.02}_{-0.02}$ & $0.48^{+0.06}_{-0.06}$ & $0.109\pm0.005$ & 0.112 \\ [0.5ex]
			12 & 563.9 & $862.05^{+0.01}_{-0.01}$ & $0.46^{+0.03}_{-0.03}$ & $0.220\pm0.007$ & 0.227 \\ [0.5ex]
			13 & 780.2 & $862.07^{+0.02}_{-0.02}$ & $0.46^{+0.05}_{-0.04}$ & $0.140\pm0.004$ & 0.145 \\ [0.5ex]
			14 & 754.3 & $862.07^{+0.01}_{-0.01}$ & $0.47^{+0.03}_{-0.03}$ & $0.199\pm0.005$ & 0.203 \\ [0.5ex]
			15 & 740.9 & $862.04^{+0.02}_{-0.02}$ & $0.49^{+0.05}_{-0.05}$ & $0.123\pm0.004$ & 0.127 \\ [0.5ex]
			16 & 719.3 & $862.09^{+0.02}_{-0.02}$ & $0.47^{+0.04}_{-0.04}$ & $0.170\pm0.007$ & 0.172 \\ [0.5ex]
			17 & 868.2 & $862.06^{+0.02}_{-0.02}$ & $0.48^{+0.07}_{-0.06}$ & $0.103\pm0.007$ & 0.111 \\ [0.5ex]
			18 & 830.6 & $862.08^{+0.02}_{-0.02}$ & $0.45^{+0.06}_{-0.05}$ & $0.100\pm0.004$ & 0.102 \\ [0.5ex]
			19 & 579.6 & $862.07^{+0.02}_{-0.02}$ & $0.48^{+0.07}_{-0.06}$ & $0.084\pm0.004$ & 0.087 \\ [0.5ex]
			20 & 944.7 & $862.06^{+0.02}_{-0.02}$ & $0.49^{+0.05}_{-0.04}$ & $0.129\pm0.004$ & 0.135 \\ [0.5ex]
			21 & 915.9 & $862.06^{+0.02}_{-0.02}$ & $0.48^{+0.06}_{-0.05}$ & $0.090\pm0.003$ & 0.094 \\ [0.5ex]
			22 & 730.9 & $862.06^{+0.01}_{-0.02}$ & $0.46^{+0.04}_{-0.04}$ & $0.124\pm0.006$ & 0.127 \\ [0.5ex]
			23 & 886.1 & $862.07^{+0.02}_{-0.02}$ & $0.50^{+0.06}_{-0.05}$ & $0.140\pm0.007$ & 0.147 \\ [0.5ex]
			24 & 589.3 & $862.04^{+0.02}_{-0.02}$ & $0.46^{+0.05}_{-0.04}$ & $0.114\pm0.004$ & 0.117 \\ [0.5ex]
			25 & 882.8 & $862.07^{+0.02}_{-0.02}$ & $0.45^{+0.07}_{-0.06}$ & $0.113\pm0.006$ & 0.116 \\ [0.5ex]
			26 & 674.8 & $862.05^{+0.02}_{-0.02}$ & $0.52^{+0.08}_{-0.07}$ & $0.100\pm0.005$ & 0.105 \\ [0.5ex]
			27 & 627.4 & $862.07^{+0.02}_{-0.02}$ & $0.47^{+0.06}_{-0.05}$ & $0.133\pm0.003$ & 0.134 \\ [0.5ex]
			28 & 491.1 & $862.06^{+0.02}_{-0.02}$ & $0.49^{+0.07}_{-0.06}$ & $0.137\pm0.005$ & 0.142 \\ [0.5ex] 
			29 & 547.3 & $862.05^{+0.02}_{-0.02}$ & $0.42^{+0.04}_{-0.04}$ & $0.102\pm0.004$ & 0.101 \\ [0.5ex] 
			\hline
			\multicolumn{6}{l}{$^a$ Signal-to-noise ratio of the red-band stacked ISM spectra.} \\
            \multicolumn{6}{l}{$^b$ Measured central wavelength in the heliocentric frame.} \\
            \multicolumn{6}{l}{$^c$ Full width at half maximum of DIB\,$\lambda$862.1.} \\
            \multicolumn{6}{l}{$^d$ Fitted equivalent width of DIB\,$\lambda$862.1.} \\
            \multicolumn{6}{l}{$^e$ Integrated equivalent width of DIB\,$\lambda$862.1.} \\
		\end{tabular}
	\end{center}
\end{table}

\begin{figure}
  \centering
  \includegraphics[width=7cm]{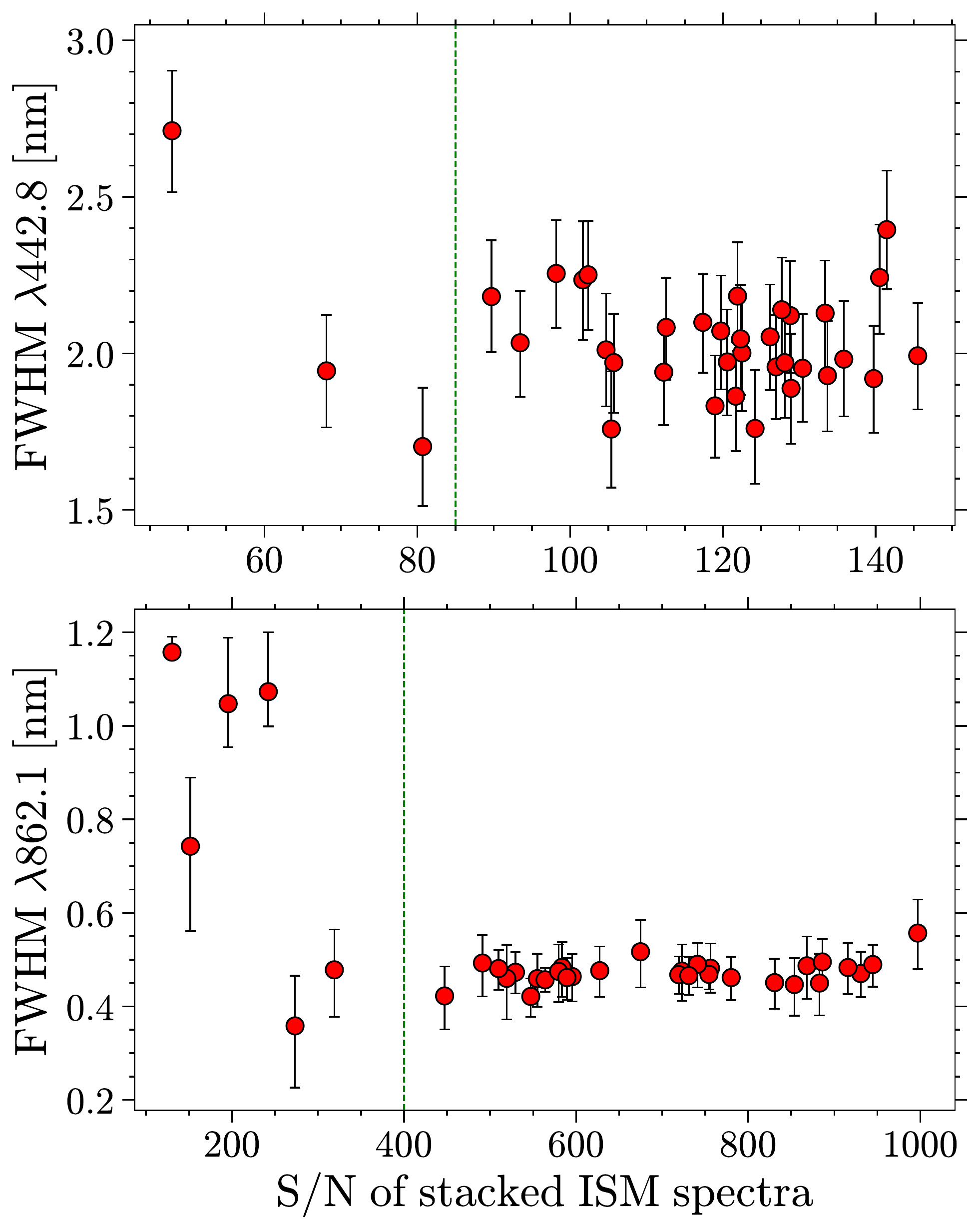}
  \caption{FHWM versus $\rm S/N$ of stacked ISM spectra for DIBs $\lambda$442.8 ({\it upper panel}) and $\lambda$862.1 ({\it lower panel}).
  The dashed green lines indicate $\rm S/N=85$ and 400, respectively.}
  \label{fig:snr-fwhm}
\end{figure}

\section{Result} \label{sect:result}

The goodness of fit is significantly affected by the $\rm S/N$ of the stacked ISM spectra. Figure \ref{fig:snr-fwhm} shows the $\rm 
S/N$ of each stacked ISM spectrum versus the corresponding fitted FWHMs of DIBs $\lambda$442.8 and $\lambda$862.1. It
can be seen that in the spectra with low $\rm S/N$, the fitted FWHM could be much larger than the average value and/or have much
larger uncertainties than average. Therefore, we limit $\rm S/N\,{>}\,85$ and $\rm S/N\,{>}\,400$ for blue-band and red-band stacked
ISM spectra, respectively, which gives us 29 fields with reliable fitting results of the two DIBs. Further analysis is based on
this sample. The stacked ISM spectra and fits to the two DIBs are shown in Figs. \ref{fig:recover-4428} and \ref{fig:recover-862.10}, 
respectively. The measured central wavelength, FWHM, and fitted equivalent width (EW) of the two DIBs $\lambda$442.8 and $\lambda$862.1 
are listed in Tables \ref{tab:stack-fit4430} and \ref{tab:stack-fit8620}, respectively. The EW uncertainty is calculated as ${\rm 
\Delta EW}\,{=}\,\sqrt{6\,{\rm FWHM}\,\delta \lambda} \times {\it R_C}$, where $\delta \lambda$ is the spectral pixel resolution 
(0.1\,nm for blue-band and 0.025\,nm for red-band) and $R_C\,{=}\,{\rm std(data-model)}$ is the noise level of the profile. This 
formula is similar to those in \citet{Vos2011} and \citet{VE2006}, who considered the main source of EW uncertainty as S/N and the 
placement of the continuum. ${\rm EW_{442.8}}$ has larger uncertainties than that of ${\rm EW_{862.1}}$ due to the lower $\rm S/N$ 
of the blue band than that of the red band.

\begin{figure}
  \centering
  \includegraphics[width=8cm]{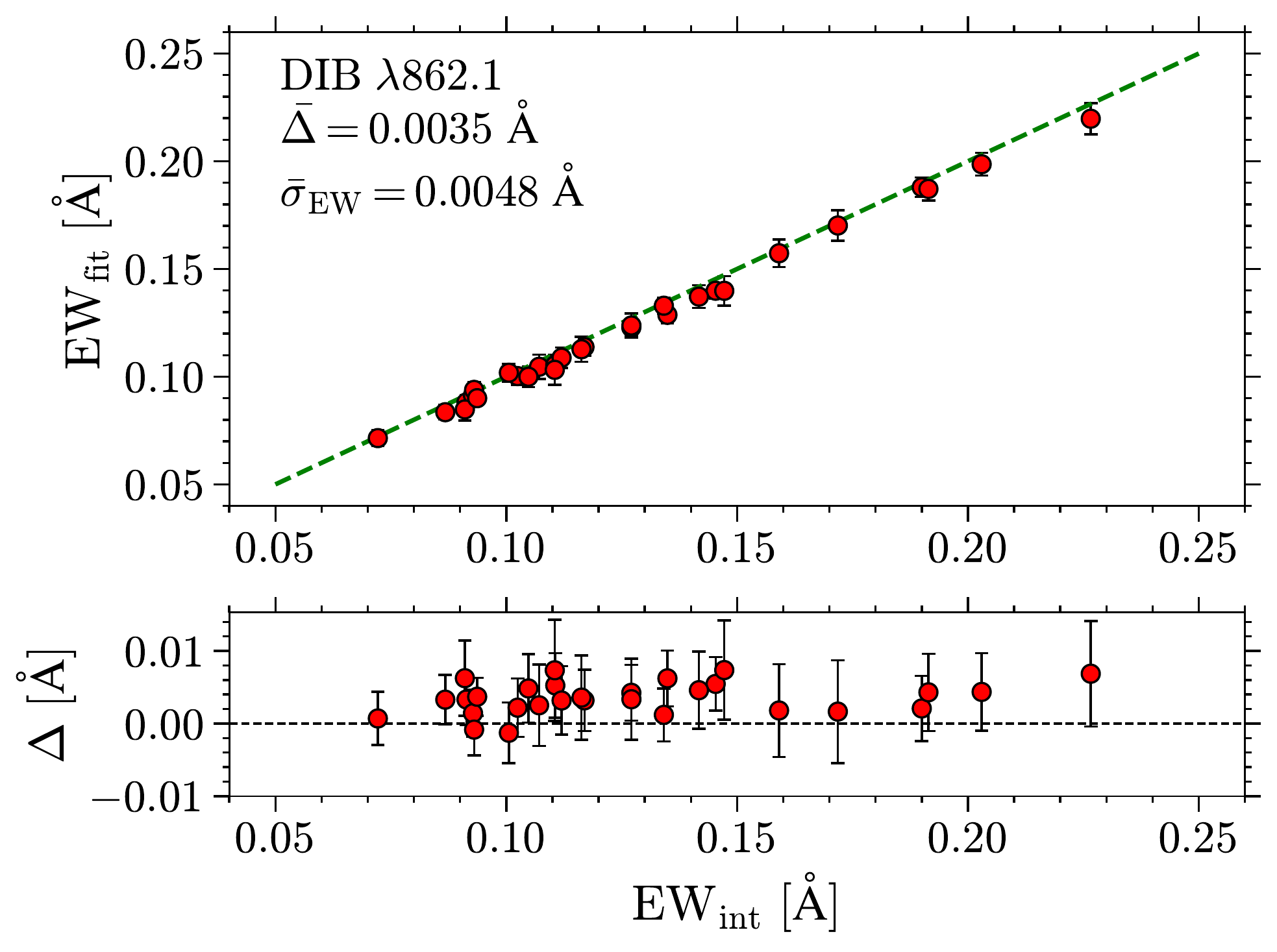}
  \caption{{\it Upper panel:} Comparison between integrated and fitted EW for DIB\,$\lambda$862.1. The dashed green line traces 
  the one-to-one correspondence. The average difference between integrated and fitted EW ($\rm \Delta=EW_{int}-EW_{fit}$) and the 
  mean EW uncertainty ($\rm \bar{\sigma}_{EW}$) are indicated. {\it Lower panel:} The EW differences as a function of the integrated
  EW. The dashed black line marks a zero difference.}
  \label{fig:int-fit}
\end{figure}

The integrated EW of $\lambda$862.1 is also calculated and listed in Table \ref{tab:stack-fit8620}. This is not done for
$\lambda$442.8 because of the stellar residuals within the DIB profile. The comparison between fitted and integrated $\rm EW_{862.1}$
is presented in Fig \ref{fig:int-fit}. The EW difference is on average smaller than its uncertainty. But the integrated EW tends
to be slightly larger than the fitted EW, which could be caused by the residuals of stellar lines near the DIB or the potential
asymmetry of the DIB profile. The fitted ${\rm EW_{442.8}}$ and ${\rm EW_{862.1}}$ are used for following analysis.


The average FWHM of $\lambda$442.8 measured in this work is $2.06\,{\pm}\,0.13$\,nm, which is slightly larger 
than the report in \citet[][1.725\,nm]{Snow2002b}. \citet{Lai2020} attributed the wide range of FWHM values of $\lambda$442.8 
in literature (e.g., 1.7\,nm in \citealt{Galazutdinov2020}; 2.4\,nm in \citealt{Fan2019}; 3.37\,nm in \citealt{Lai2020}) to the
differences in the local radiation field. The average FWHM of $\lambda$862.1 measured in this work is $0.47\,{\pm}\,0.03$\,nm, 
which is close to the measure of 0.43\,nm in \citet{HL1991} and 0.469\,nm in \citet{Maiz-Apellaniz2015a}.

\subsection{Latitude groups} \label{subsect:group}

\begin{figure}
  \centering
  \includegraphics[width=8cm]{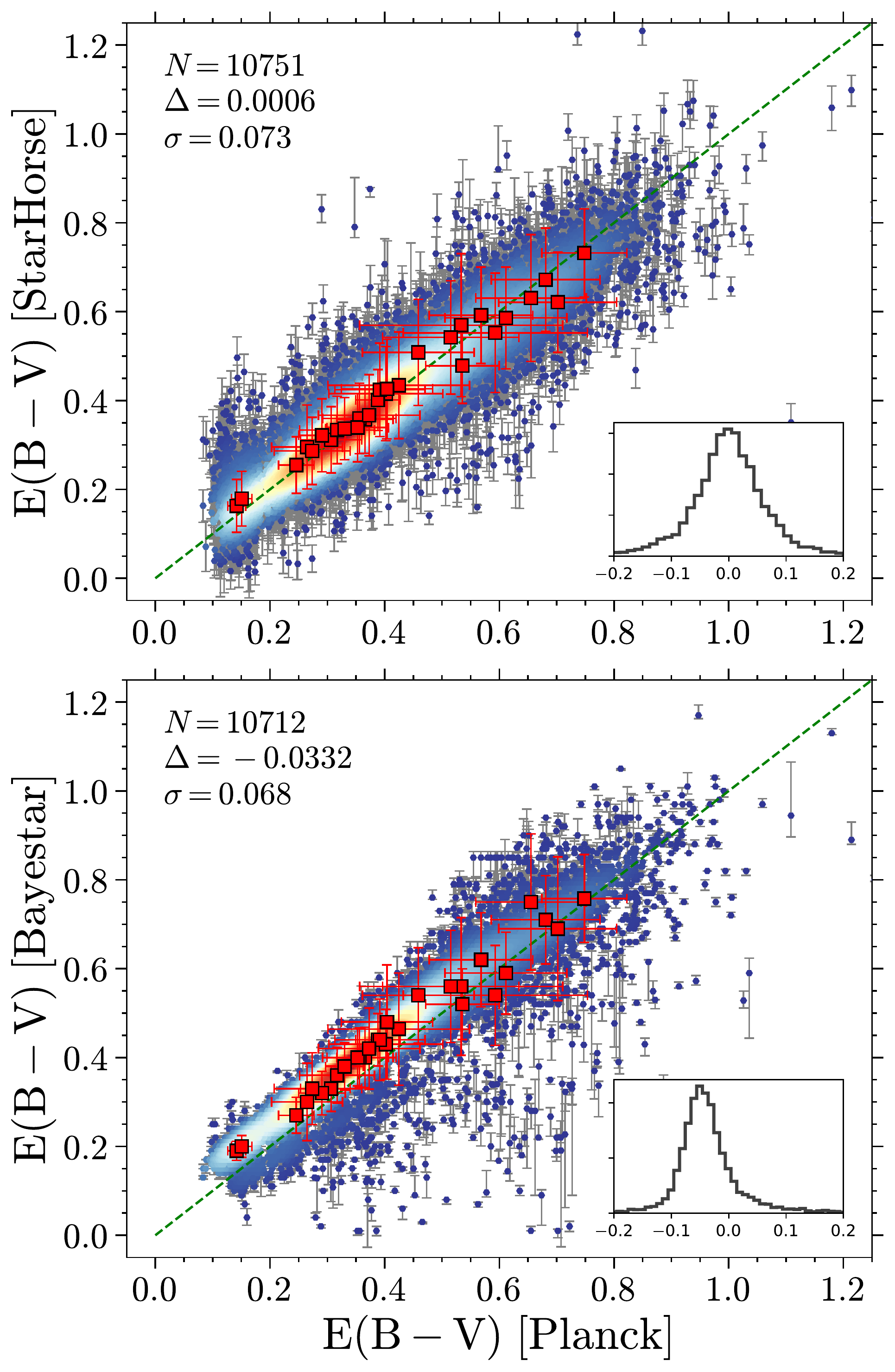}
  \caption{Comparison of $\EBV$ derived from \citet{Planck2016dust} with those from the StarHorse algorithm ({\it upper panel}) and
  from the Bayestar map ({\it lower panel}). The colored dots with uncertainties are for individual PIGS target stars. Their colors
  represent the number densities estimated by a Gaussian kernel density estimation. The zoom-in panels show the $\EBV$ differences
  of Planck--StarHorse and Planck--Bayestar, respectively. The number of PIGS target stars ($N$), the mean differences ($\Delta$), 
  and its standard deviation ($\sigma$) are indicated in the panels. The red squares with errorbars are the median $\EBV$ and their 
  standard deviations calculate in 29 used PIGS fields. The dashed green lines trace the one-to-one correspondence.}
  \label{fig:ebv}
\end{figure}

In this work, we derive $\EBV$ for each PIGS target from the map of \citet{Planck2016dust} using the python package {\it dustmaps}
\citep{Green2018python} because these target stars are mainly very distant and at high latitudes ($93.6\%$ with $|b|\,{>}\,4^{\circ}$). 
The median $\EBV$ in each PIGS field, together with its standard deviation as a measure of uncertainty, are listed in Table 
\ref{tab:stack-fit4430}. To check the reliability of the Planck map for PIGS targets, we compare the reddening values with 
estimates from two other sources: the 3D reddening values derived with the StarHorse algorithm \citep{Queiroz2018} specifically for 
the PIGS stars using the PIGS spectroscopic stellar parameters, Pan--STARRS1 photometry, and {\it Gaia} parallaxes 
(Arentsen et al. in prep.), and the 3D Bayestar reddening map \citep{Green2019} applying the StarHorse distances. The usage of PIGS 
stellar parameters into StarHorse delivers a more constrained and less uncertain reddening and distances than those from the StarHorse 
{\it Gaia} database \citep{Anders2022} for the same stars, but that only used photometry and parallaxes as input. About 80\% PIGS 
target stars have StarHorse and Bayestar $\EBV$, among which, 90\% stars are further than 5\,kpc. The comparison between the Planck, 
StarHorse, and Bayestar reddenings is presented in Fig. \ref{fig:ebv}. $\EBV$ from Planck and StarHorse are highly consistent with 
each other (the mean difference is smaller than one thousandth magnitude), while the Bayestar $\EBV$ is slightly larger (0.033\,mag 
on average) than the Planck values, which could be due to the different methods of reddening inference. The differences of median 
$\EBV$ in the 29 used PIGS fields (red squares in Fig. \ref{fig:ebv}) between Planck, StarHorse, and Bayestar are mostly smaller 
than their uncertainties. This ensures that the usage of Planck $\EBV$, which can be derived for all the PIGS targets, can be a 
safe measure of the dust column densities toward these sightlines.

\begin{figure}
  \centering
  \includegraphics[width=8cm]{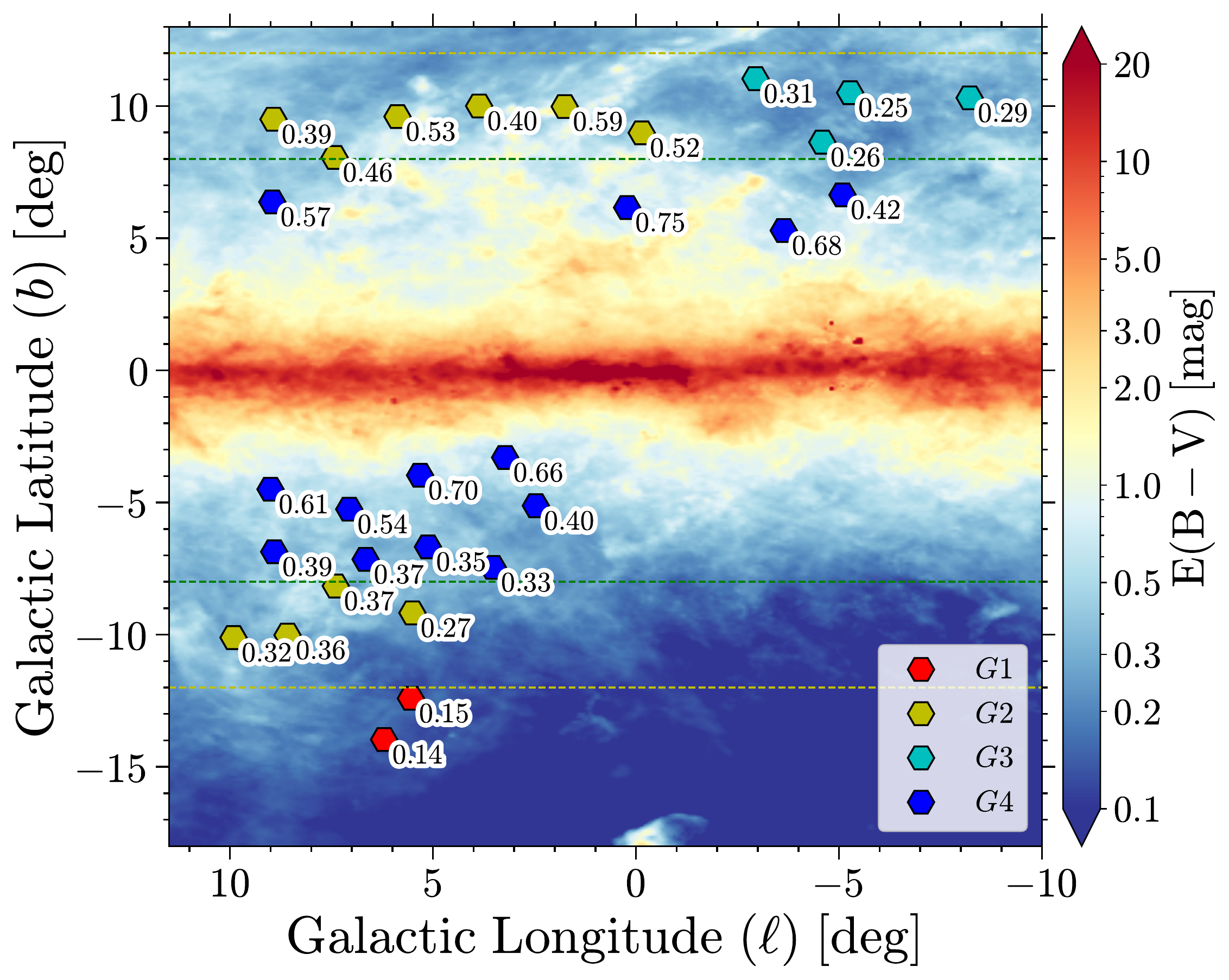}
  \caption{Spatial distribution $(\ell,b)$ of 29 PIGS fields, overplotted on the dust reddening map of \citet{Planck2016dust}. 
    Different colors indicate different latitude groups defined in Sect. \ref{subsect:group}, that is red: G1, $|b|\,{>}\,12^{\circ}$; 
    yellow: G2, $8^{\circ}\,{<}\,|b|\,{<}\,12^{\circ}$ and $\ell\,{>}\,{-}1^{\circ}$; cyan: G3, $8^{\circ}\,{<}\,|b|\,{<}\,12^{\circ}$ 
    and $\ell\,{<}\,{-}1^{\circ}$, and blue: G4, $|b|\,{<}\,8^{\circ}$. The median $\EBV$ of each filed is also indicated.}
  \label{fig:bgroup}
\end{figure}

In our sample, $\EBV$ is strongly correlated with the Galactic latitude. Thus, the PIGS fields are divided into three latitude 
stripes to highlight the effect of latitude in the following analysis. The middle stripe is further separated into two at 
$\ell\,{=}\,{-}1^{\circ}$ considering the effect of $\ell$ toward the Galactic center (GC). Finally, we roughly define four 
latitude groups (see Fig. \ref{fig:bgroup}): G1: $|b|\,{>}\,12^{\circ}$ (red), G2: $8^{\circ}\,{<}\,|b|\,{<}\,12^{\circ}$ and 
$\ell\,{>}\,{-}1^{\circ}$ (yellow), G3: $8^{\circ}\,{<}\,|b|\,{<}\,12^{\circ}$ and $\ell\,{<}\,{-}1^{\circ}$ (cyan), and G4: 
$|b|\,{<}\,8^{\circ}$ (blue).

\subsection{Linear relations between different interstellar materials} \label{subsect:linear}

\begin{figure}
  \centering
  \includegraphics[width=7cm]{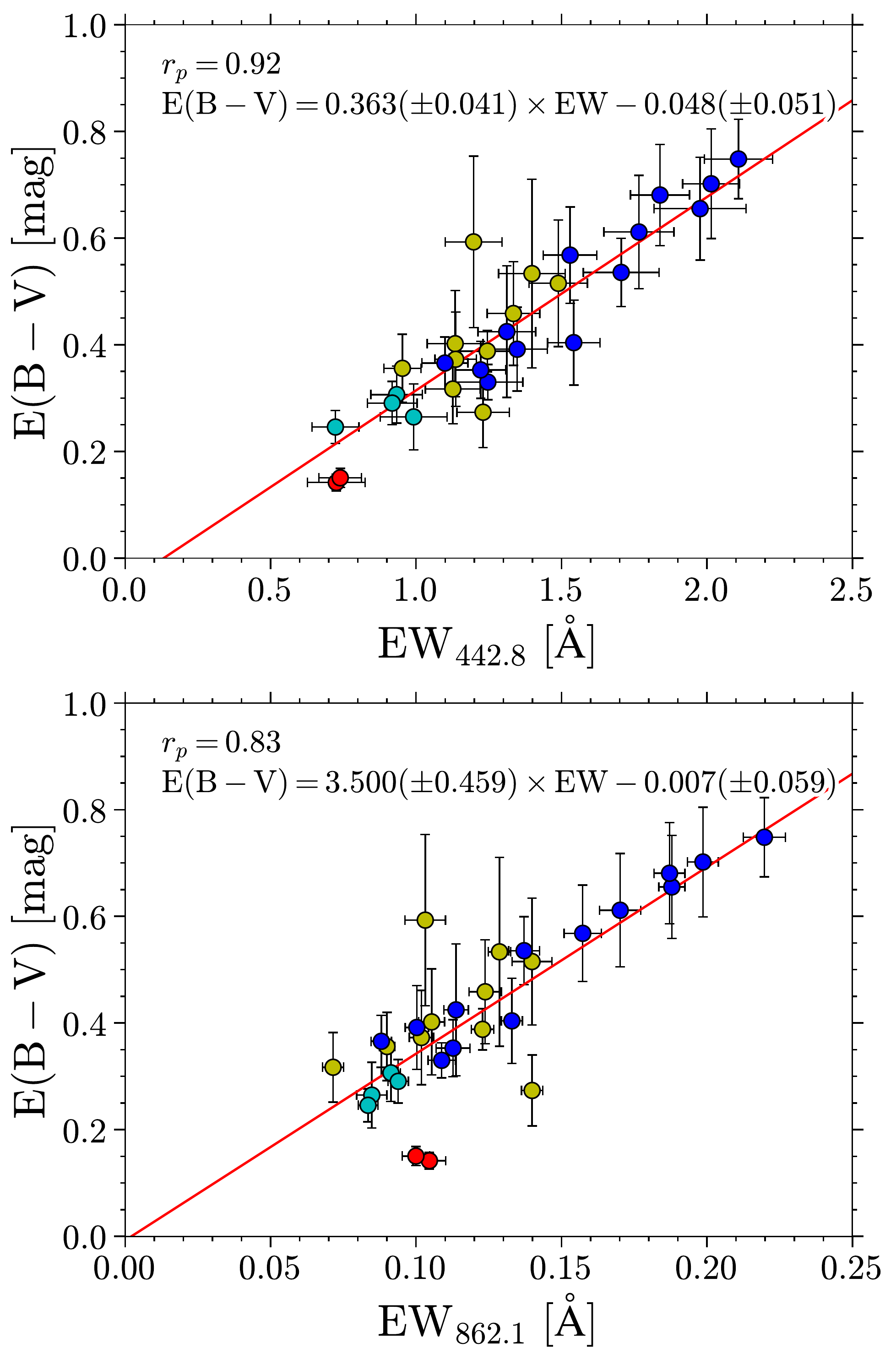}
  \caption{Correlation between DIB EW of $\lambda$442.8 ({\it upper panel}) and $\lambda$862.1 ({\it lower panel}) measured in stacked 
  ISM spectra and median $\EBV$ from \citetalias{Planck2016dust} map in corresponding fields. The red lines are the linear fits.  
  The fitting results and the Pearson coefficient ($r_p$) are also indicated. The points in different colors belong to different
  latitude groups defined in Sect. \ref{subsect:group} and shown in Fig. \ref{fig:bgroup}.}
  \label{fig:ew-ebv}
\end{figure}

One of the basic characteristics of most strong DIBs is the increase of their strength with dust reddenings. Figure \ref{fig:ew-ebv} 
shows the correlation between DIB strength ($\rm EW_{442.8}$ and $\rm EW_{862.1}$) and $\EBV$. Linear correlation with $\EBV$ can 
be found for both $\lambda$442.8 and $\lambda$862.1 with a Pearson coefficient of $r_p\,{=}\,0.92$ and $r_p\,{=}\,0.83$, respectively. 
The two outliers in G2 (yellow points) are due to the local variation of $\EBV$, that one with $\EBV\,{=}\,0.59$\,mag is higher 
than its vicinity and the other with $\EBV\,{=}\,0.27$\,mag is lower than its neighboring values (see Fig. \ref{fig:bgroup}). 
Deviations from the linear correlation between DIB and $\EBV$ can also be found at high
latitudes (see red points in Fig. \ref{fig:ew-ebv}), which will be discussed in detail below.
A linear fit of $\EBV=0.363(\pm0.041) \times {\rm EW_{442.8}} - 0.048(\pm0.051)$ corresponds to ${\rm EW_{442.8}}/\EBV\,{=}\,2.75$\,{\AA}\,mag$^{-1}$,
which is in the intermediate range compared to the results in literature (e.g., 2.89 of \citealt{Isobe1986} and 2.01 of \citealt{Fan2019}). For
$\lambda$862.1, we derived a coefficient of $\EBV/{\rm EW_{862.1}}\,{=}\,3.500\pm0.459$\,mag\,{\AA}$^{-1}$ with a very small intercept
(${-}0.007\pm0.059$). This value is slightly larger than previous results \citep[e.g.,][]{Munari2008,Kos2013,Krelowski2019b} but 
between the values derived from {\it Gaia}--DR3 DIB results with different $\EBV$ sources \citep[see Table 3 in][]{Schultheis2022}. 

It has been known that the ratio of $\EBV/{\rm EW}$ can vary significantly from one sightline to another and is also affected by
the use of different data samples and methods. Nevertheless, we argue that the positive correlation between DIB strength and dust 
reddening in diffuse or intermediate ISM with a proper coefficient can be treated as a validation of the DIB measurement. For our 
results, the range of $\rm EW_{442.8}$ at given $\EBV$ between 0.2 and 0.8\,mag is consistent with archival data shown in Fig. 5 
in \citet{Lai2020} and early results of \citet{Herbig1975} shown in Fig. 6 in \citet{Snow2002b}. The variation of $\rm EW_{862.1}$ 
relative to $\EBV$ is also within the regions, considering the scatter, shown in Fig. 8 in \citet{Schultheis2022}. 

\begin{figure}
  \centering
  \includegraphics[width=7cm]{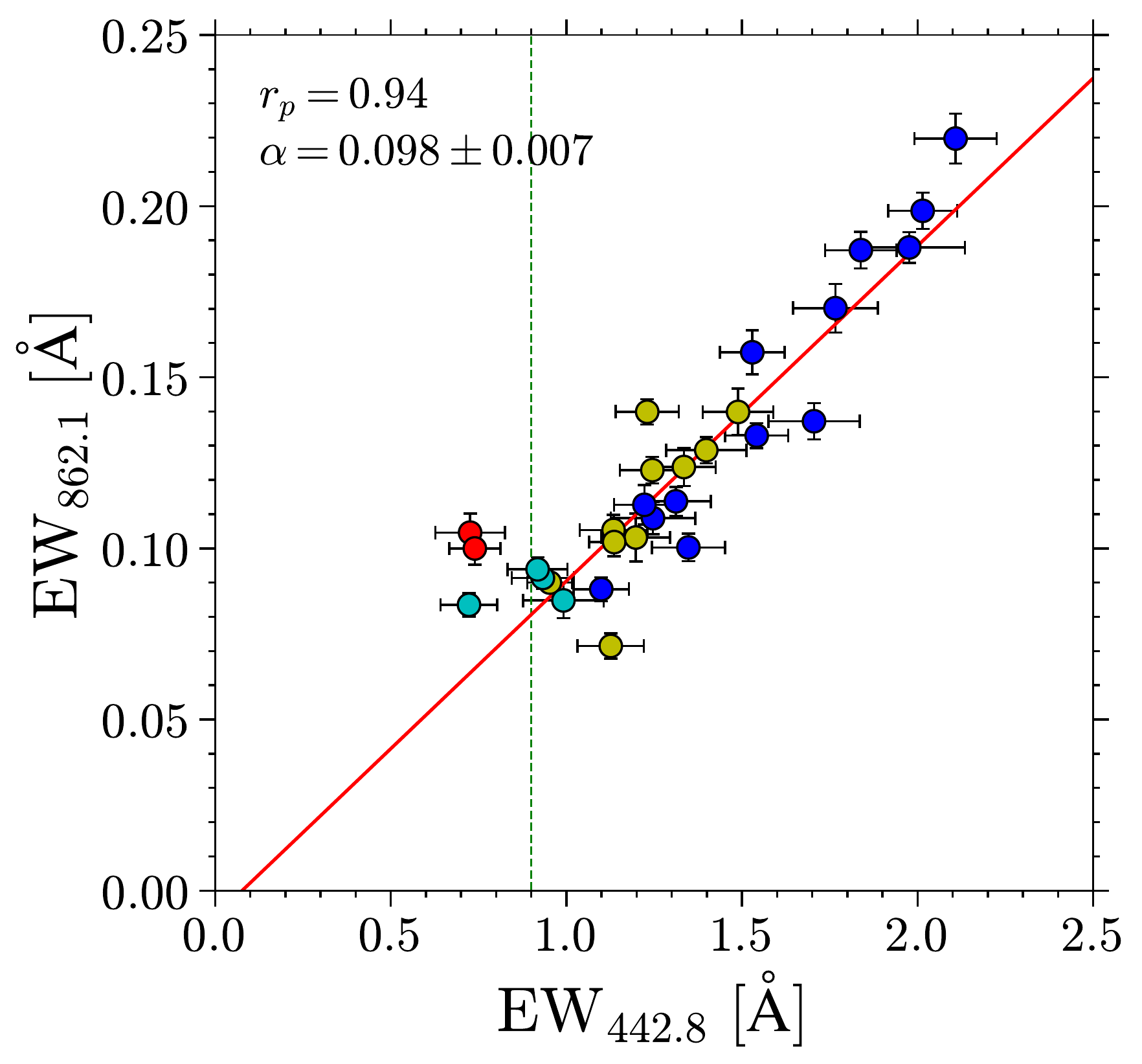}
  \caption{Correlation between $\rm EW_{442.8}$ and $\rm EW_{862.1}$ measured in blue-band and red-band stacked ISM spectra, 
  respectively. The red line is the linear fit to the dots with $\rm EW_{442.8}>0.9$\,{\AA} (indicated by the dashed green 
  line). The fitted slope ($\alpha$) and Pearson coefficient ($r_p$) are indicated. The points are colored differently 
  according to the latitude groups they are assigned to (see Sect. \ref{subsect:group} and Fig. \ref{fig:bgroup}).}
  \label{fig:dib-dib}
\end{figure}

The tight linear correlation ($r_p\,{=}\,0.94$) between the DIB strength of $\lambda$442.8 and $\lambda$862.1 can be seen in Fig. 
\ref{fig:dib-dib} for $\rm EW_{442.8}\,{>}\,0.9$\,{\AA}. A linear fit yields $\rm EW_{862.1}/EW_{442.8}=0.098\pm0.007$ with a very
small offset of ${-}0.008\pm0.009$. Note that the strongest DIB $\lambda$442.8 is stronger than $\lambda$862.1 by a factor of 10
for our results, which is consistent with the average of their relative strength measured in \citet{Fan2019}.
However, three fields with $\rm EW_{442.8}\,{<}\,0.9$\,{\AA} have $\rm EW_{862.1}$ much larger than that expected by the linear relation. 

\subsection{Variation of the relative strength with Galactic latitude and reddening} \label{subsect:variation}

\begin{figure}
  \centering
  \includegraphics[width=7cm]{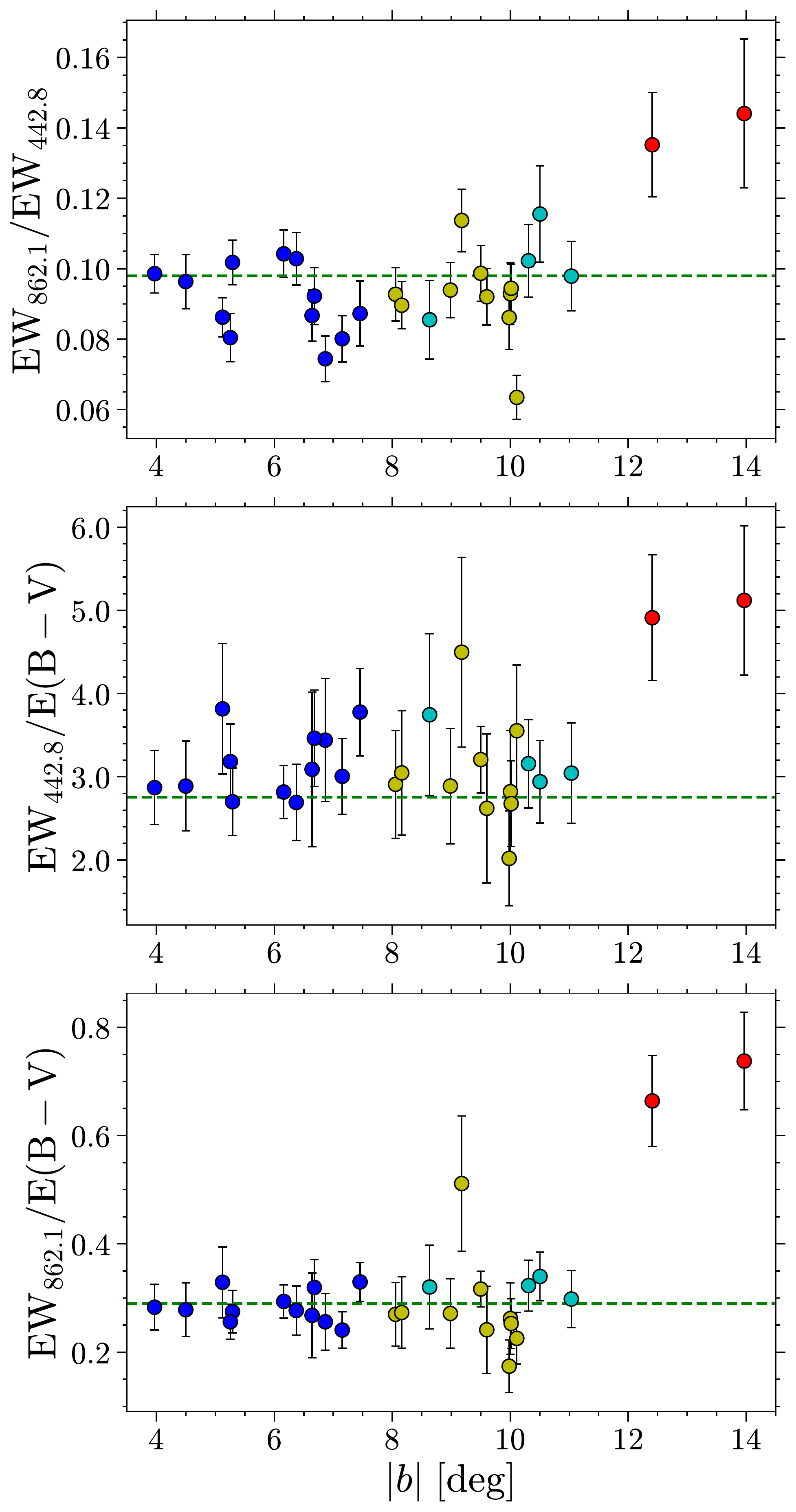}
  \caption{Variation of relative strength between DIBs $\lambda$442.8, $\lambda$862.1, and dust with Galactic latitude ($|b|$). The 
  dashed green lines indicate their average strength ratios from linear fits (see Sect. \ref{subsect:linear}). See Sect. 
  \ref{subsect:group} and Fig. \ref{fig:bgroup} for the point colors representing different latitude groups.}
  \label{fig:var-b}
\end{figure}

\begin{figure}
  \centering
  \includegraphics[width=7cm]{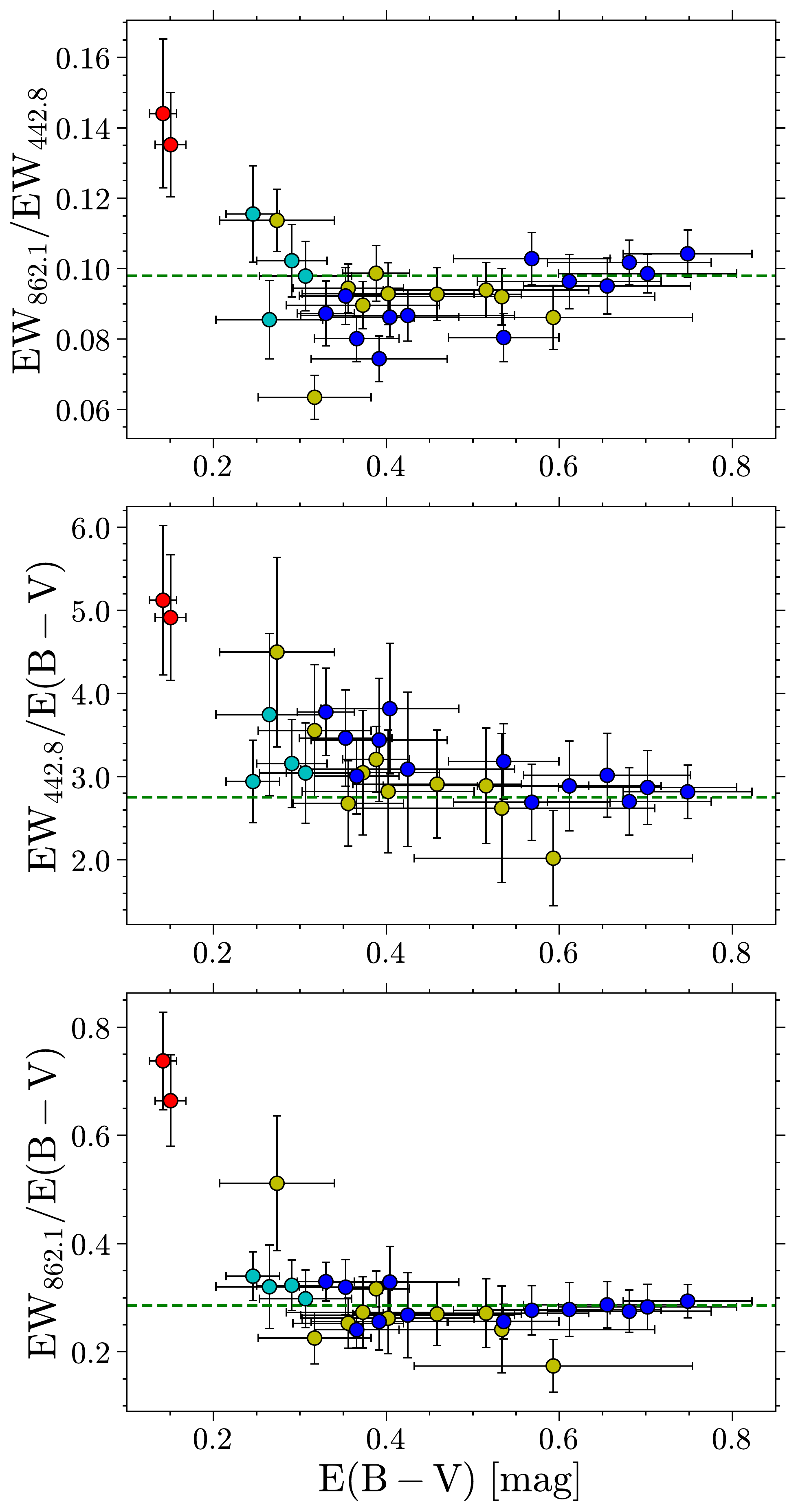}
  \caption{The same as Fig. \ref{fig:var-b}, but for the variation with $\EBV$.}
  \label{fig:var-ebv}
\end{figure}

The systematic variation of $\rm EW_{862.1}/EW_{442.8}$ with the Galactic latitude ($|b|$) and dust reddening ($\EBV$) are presented in
the upper panels in Figs. \ref{fig:var-b} and \ref{fig:var-ebv}, respectively, whose uncertainty considers both the contribution 
of $\rm EW_{862.1}$ and $\rm EW_{442.8}$ by error propagation. $\rm EW_{862.1}/EW_{442.8}$ becomes larger than average for 
$|b|\,{\gtrsim}\,10^{\circ}$ or $\EBV\,{\lesssim}\,0.3$\,mag, where an increase of $\rm EW_{862.1}/EW_{442.8}$ can be found with the 
increasing $|b|$ and the decreasing $\EBV$. The uncertainty of $\rm EW_{862.1}/EW_{442.8}$ tends to be larger in the fields with high 
latitudes or small $\EBV$, which have a risk to blur the variation of the DIB relative strength. But we notice that for the G1 fields 
(red points in Figs. \ref{fig:var-b} and \ref{fig:var-ebv}) with $\EBV\,{<}\,0.2$\,mag and $|b|\,{>}\,12^{\circ}$, the increasing 
magnitude of their mean $\rm EW_{862.1}/EW_{442.8}$ 0.140) to the fitted coefficient (0.098) is 0.042, which is bigger than their 
mean uncertainty (0.018) by a factor of two. Moreover, a tight negative correlation ($r_p\,{=}\,{-}0.88$) can be found between 
$\rm EW_{862.1}/EW_{442.8}$ and $\EBV$ for $\EBV\,{<}\,0.31$\,mag, which also confirms that the variation of $\rm EW_{862.1}/EW_{442.8}$
with $|b|$ and $\EBV$ is not caused by the EW uncertainty but indicates the different distributions of the carriers of the two DIBs
(see Sect. \ref{sect:discuss} for more discussions). For $\EBV\,{\gtrsim}\,0.45$\,mag, $\rm EW_{862.1}/EW_{442.8}$
tends to slightly increase with $\EBV$ (see top panel in Fig. \ref{fig:var-ebv}), but more data are needed to confirm this trend. 
We also emphasize that in Fig. \ref{fig:dib-dib}, three G3 fields (cyan points) 
with $\rm EW_{442.8}\,{>}\,0.9$\,{\AA} were used for the linear fit of DIB strength to make the offset of the line close to zero,
but their $\rm EW_{862.1}/EW_{442.8}$ already present a systematic variation with respect to $\EBV$. 

In our sample, fields at high latitudes in general have small $\EBV$, but the dust distribution shown in Fig \ref{fig:bgroup} also 
varies with longitudes and sightlines. Consequently, we can find a G3 field, with $(\ell,b)\,{=}\,({-}4.59^{\circ},8.63^{\circ})$ 
and $\EBV\,{=}\,0.26$\,mag, that follows the negative trend between $\rm EW_{862.1}/EW_{442.8}$ and $\EBV$ but does not present a 
clear variation of $\rm EW_{862.1}/EW_{442.8}$ with $|b|$. The non-monotonic relationship between $|b|$ and $\EBV$, as well as the
averaging of ISM and the complicated environments toward the GC, also introduces the scatters in Figs. \ref{fig:var-b} and 
\ref{fig:var-ebv}. 

Similar pictures can also be found for ${\rm EW_{442.8}}/\EBV$ and ${\rm EW_{862.1}}/\EBV$ which are shown in the middle and lower 
panels in Figs. \ref{fig:var-b} and \ref{fig:var-ebv}, respectively. A remarkable increase of ${\rm EW}/\EBV$ can be found 
for two G1 fields in our sample for both $\lambda$442.8 and $\lambda$862.1. ${\rm EW_{862.1}}/\EBV$ stay around the 
linear relation in a wide range of $|b|\,{\lesssim}\,11^{\circ}$ or $\EBV\,{\gtrsim}\,0.3$\,mag. Nevertheless, ${\rm EW_{442.8}}/\EBV$ 
presents larger scatters with respect to $|b|$ and $\EBV$ than ${\rm EW_{862.1}}/\EBV$, which implies that the carrier abundance 
of $\lambda$442.8 would be more sensitive to the dust column densities and latitudes than that of $\lambda$862.1. ${\rm EW}/\EBV$ 
tends to be larger than average for $\EBV$ between 0.25 and 0.4\,mag, but no linear tendency can be found.


\section{Discussion} \label{sect:discuss}

\subsection{Relative vertical distributions between DIB carriers and dust grains} \label{subsect:distribution1}

By covering a wide range of Galactic latitude ($4\degr\,{<}\,|b|\,{<}\,15\degr$) and dust reddening ($0.1\,{<}\,\EBV\,{<}\,0.8$\,mag), 
our results show that the DIBs $\lambda$442.8 and $\lambda$862.1 engage similar behavior with dust grains, that is the change of 
their ${\rm EW}/\EBV$ is constant and around the mean value considering the uncertainties for $|b|\,{<}\,12^{\circ}$ or 
$\EBV\,{>}\,0.3$\,mag, which indicates that the abundance of the DIB carriers and dust grains increase with each other
in the Galactic middle plane. On the other hand, ${\rm EW}/\EBV$ becomes significantly larger than average in the G1
fields at high latitudes. This phenomenon could be interpreted as an evidence that DIBs and dust grains have different vertical 
distributions in the Milky Way, because in our sample small $\EBV$ generally indicate sight lines towards higher latitudes. 
However, clearly more data especially at higher galactic latitudes are necessary to confirm the different vertical distributions 
between the DIBs and the dust.

For our present sample, we cannot quantitatively estimate a scale height for the DIB carrier or dust because of the limited
sample size and the gaps at $|b|\,{\sim}\,12^{\circ}$ and $\EBV\,{\sim}\,0.2\,$mag. But the increase of ${\rm EW}/\EBV$ at high 
latitudes indicates the decrease of the column density of dust grains with respect to that of the DIB carriers, implying a larger 
scale height of the DIB carriers for both $\lambda$442.8 and $\lambda$862.1 than that of the dust grains if we assume a
simple disk model for them and that one with a larger scale height would have a larger scale length as well. This result is consistent 
with the detection of DIBs toward sightlines with negligible reddenings at high latitudes \citep{Baron2015b} and the result of 
\citet{Kos2014} who measured $\lambda$862.1 in RAVE spectra in a range of $240^{\circ}\,{\leqslant}\,\ell\,{\leqslant}\,330^{\circ}$.
However, based on the {\it Gaia}--DR3 results, \citet{Schultheis2022} determined a scale height of DIB\,$\lambda$862.1 as
$98.69^{{+}10.81}_{{-}8.35}$\,pc in a range of $0^{\circ}\,{\leqslant}\,{\ell}\,{\leqslant}\,360^{\circ}$ and 
$4^{\circ}\,{\leqslant}\,{|b|}\,{\leqslant}\,12^{\circ}$, which is smaller than the usually suggested scale height of dust grains, 
such as $134.4{\pm}8.5$\,pc \citep{DS2001} and $125^{{+}17}_{{-}7}$\,pc \citep{Marshall2006}. The discrepancy could be a result of
the variation of the distribution of DIB carriers and dust grains from one sightline to another (see the wavy pattern of the dust
shown in \citealt{Lallement2022} for example). Furthermore, the vertical distribution of interstellar materials would be more
complicated than a single exponential model. \citet{GuoHL2021} fitted the dust distribution with a two-disk model and got two scale 
heights of $72.7{\pm}2.4$\,pc and $224.6{\pm}0.7$\,pc for the thin and thick disks. Similarly, \citet{Su2021} also characterized the 
molecular disk, traced by $^{12}$CO $J\,{=}\,(1{-}0)$ emission \citep{Su2019}, by two components with a thickness of ${\sim}85$\,pc 
and ${\sim}280$\,pc, respectively, in a range of $16^{\circ}\,{\leqslant}\,{\ell}\,{\leqslant}\,25^{\circ}$ and ${|b|}\,{<}\,5.1^{\circ}$. 
The {\it Gaia} result presents an average of the scale height of the carrier of $\lambda$862.1 in vicinity of the Sun ($\lesssim$3\,kpc). 
Nevertheless, the PIGS target stars are located toward the GC ($\ell\,{<}\,11^{\circ}$) and more distant (90\% $>$5\,kpc) that could 
trace a different relative vertical distribution between DIB carriers and dust grains, if we expect their distributions vary in
different manners.


\subsection{Different vertical distributions between carriers of different DIBs} \label{subsect:distribution2}

Tight intensity correlations have been reported for many strong optical DIBs \citep[e.g.,][]{Friedman2011,Xiang2012,KZ2013}. But
most of these works rely on OB stars that mainly reside in the Galactic middle plane, where one can always get a linear relationship 
between different interstellar materials in a broad enough distance range. Thus, a tight linearity is a necessary but not sufficient 
condition to conclude a common origin for different DIBs. An example is $\lambda$578.8 and $\lambda$579.7 that they have been proved 
to have different origins \citep[e.g.,][]{KW1988,Cami1997,KZ2013} but high-level correlations can still be found with $r_p\,{>}\,0.9$ 
\citep[e.g.,][]{Friedman2011,Xiang2012,KZ2013}. 

The variation of $\rm EW_{862.1}/EW_{442.8}$ with $|b|$ and $\EBV$ is a strong evidence that $\lambda$442.8 and $\lambda$862.1 do 
not share a common carrier. Moreover, their carriers could be well mixed in the fields with high reddenings or low latitudes seen
though their tight intensity correlation ($r_p\,{=}\,0.94$). But the carrier of $\lambda$862.1 becomes more abundant with respect 
to $\lambda$442.8 at higher latitudes, which is consistent with \citet{MW1993} who found that the strength of $\lambda$442.8 
relative to those of $\lambda$578.0 and $\lambda$5797.7 was greatest at low latitude and decreased with increasing latitude. 
\citet{Baron2015a} also showed that $\lambda$442.8 was absent in their spectra at high latitudes while $\lambda$578.0 and $\lambda$5797.7 tended to 
have higher EW per reddening. Our results are in agreement with their findings that different DIB carriers could present different 
vertical distributions and the carrier of $\lambda$442.8 seems to be mainly located in the Galactic plane compared to other DIBs.
A different origin for $\lambda$442.8 and $\lambda$862.1 is not unexpected as they are far away from each other in wavelength
and their profiles show different shapes. However, they can be treated as an example to illustrate the significance 
to explore the spatial distributions, especially for high-latitude regions, of different DIBs when we would like to confirm a 
common or different origin for them. 

A potential risk of the above interpretation is the change of the environmental conditions, like temperature, from the 
Galactic middle plane to the regions far away from it. It is therefore possible that a single DIB carrier produces two DIBs from 
two different transitions that show different vertical structures. As $C_{60}^{+}$ is the only identified DIB carrier and the 
variation of its DIBs with environments has not been observed, it is hard to characterize this effect. Further studies, combined
with other known atomic or molecular species, could address this problem to some extent. In the future, much can be gained from 
the large sky-area spectroscopic surveys to investigate if there exists a layered structure along the vertical direction for a set 
of DIBs, revealing a hierarchical distribution of various macromolecules or a dependence of their electronic transitions
on the interstellar environments.

\section{Conclusions} \label{sect:conclusion}

Based on stacking blue-band and red-band ISM spectra from the PIGS sample, we successfully fitted and measured two DIBs 
$\lambda$442.8 and $\lambda$862.1 in 29 fields with a mean radii of 1{\degr}. Their FWHM was estimated as 
$2.06\,{\pm}\,0.13$\,nm for $\lambda$442.8 and $0.47\,{\pm}\,0.03$\,nm for $\lambda$862.1, which are both consistent with previous 
measurements.

Our results depict a general image of the relative distributions of two DIBs and dust grains toward the GC with $|\ell|\,{<}\,11^{\circ}$ 
and $4\degr\,{<}\,|b|\,{<}\,15\degr$. The DIB carriers and dust grains are well mixed with each other for $|b|\,{<}\,12^{\circ}$ 
or $\EBV\,{>}\,0.3$\,mag. Tight linear correlations are derived between EW and $\EBV$ for both $\lambda$442.8 ($r_p\,{=}\,0.92$) 
and $\lambda$862.1 ($r_p\,{=}\,0.83$). For $|b|\,{>}\,12^{\circ}$, $\lambda$442.8 and $\lambda$862.1 have larger relative 
strength with respect to the dust grains, which implies a larger scale heights of the carriers of $\lambda$442.8 and $\lambda$862.1 
than that of dust grains toward the GC.

A tight linear intensity correlation ($r_p\,{=}\,0.94$) is also found between $\lambda$442.8 and $\lambda$862.1 when $|b|\,{\lesssim}\,10^{\circ}$ 
or $\EBV\,{\gtrsim}\,0.3$\,mag, with a relative strength of $\rm EW_{862.1}/EW_{442.8}=0.098\pm0.007$. But an increase of $\rm EW_{862.1}/EW_{442.8}$ 
with the increasing $|b|$ and the decreasing $\EBV$ for the fields at high latitudes concludes different carriers for the two DIBs. 
Our results suggest that the carrier of $\lambda$862.1 could have a larger scale height than that of $\lambda$442.8.

This work can be treated as an example to show the significance and potentials of the DIB research covering a large range of 
latitudes listed below:

\begin{enumerate}
    \item The variation of the DIB relative strength at high latitudes is a strong evidence to conclude a common or different 
    origin for different DIBs.
    \item Vertical distributions of different DIBs can help us to reveal the structure of the Galactic ISM, especially the
    carbon-bearing macromolecules which are supposed to be the DIB carriers.
    \item Relative distributions between different DIBs are also clues of their carrier properties. For example, a DIB with larger
    scale height would imply that its carrier can be formed earlier or more quickly in the Galactic plane and then be transported
    to the high-latitude regions. Alternatively, we would trace carriers formed in the Galactic halo.
\end{enumerate}

\section*{Acknowledgements}

HZ is funded by the China Scholarship Council (No. 201806040200). AA and NFM acknowledge funding from the European Research Council 
(ERC) under the European Unions Horizon 2020 research and innovation programme (grant agreement No. 834148). ES acknowledges funding 
through VIDI grant ``Pushing Galactic Archaeology to its limits'' (with project number VI.Vidi.193.093) which is funded by the Dutch 
Research Council (NWO). MS, VH, and NFM gratefully acknowledge support from the French National Research Agency (ANR) funded project 
``Pristine'' (ANR-18-CE31-0017). 

\section*{Data Availability}

The DIB fitting results in each fields are shown in Tables \ref{tab:stack-fit4430} and \ref{tab:stack-fit8620}.
The spectra underlying this article will be shared on reasonable request to Anke Arentsen.



\bibliographystyle{mnras}
\bibliography{hzref} 




\appendix

\section{DIB fitting in stacked ISM spectra} \label{appsec:fit-stack}

\begin{figure*}
    \centering
    \includegraphics[width=16cm]{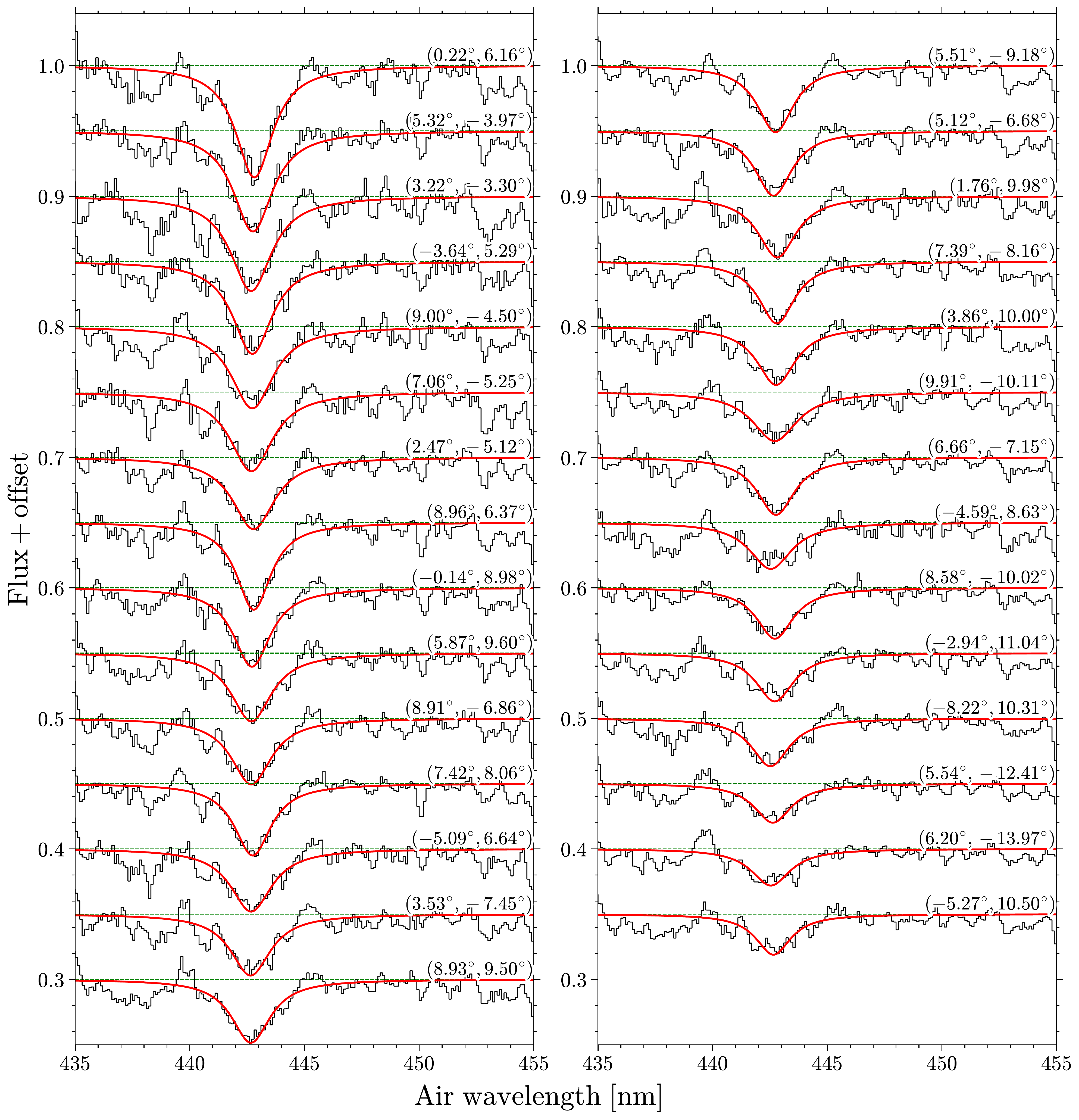}
    \caption{Fits of $\lambda$442.8 in blue-band stacked ISM spectra. The black lines are the stacked ISM spectra, and the red lines
    are the fitted DIB profile. The results are sorted by the measured $\rm EW_{442.8}$. The Galactic coordinates $(\ell,b)$ of each
    field are also indicated.}
    \label{fig:recover-4428}
\end{figure*}

\begin{figure*}
    \centering
    \includegraphics[width=16cm]{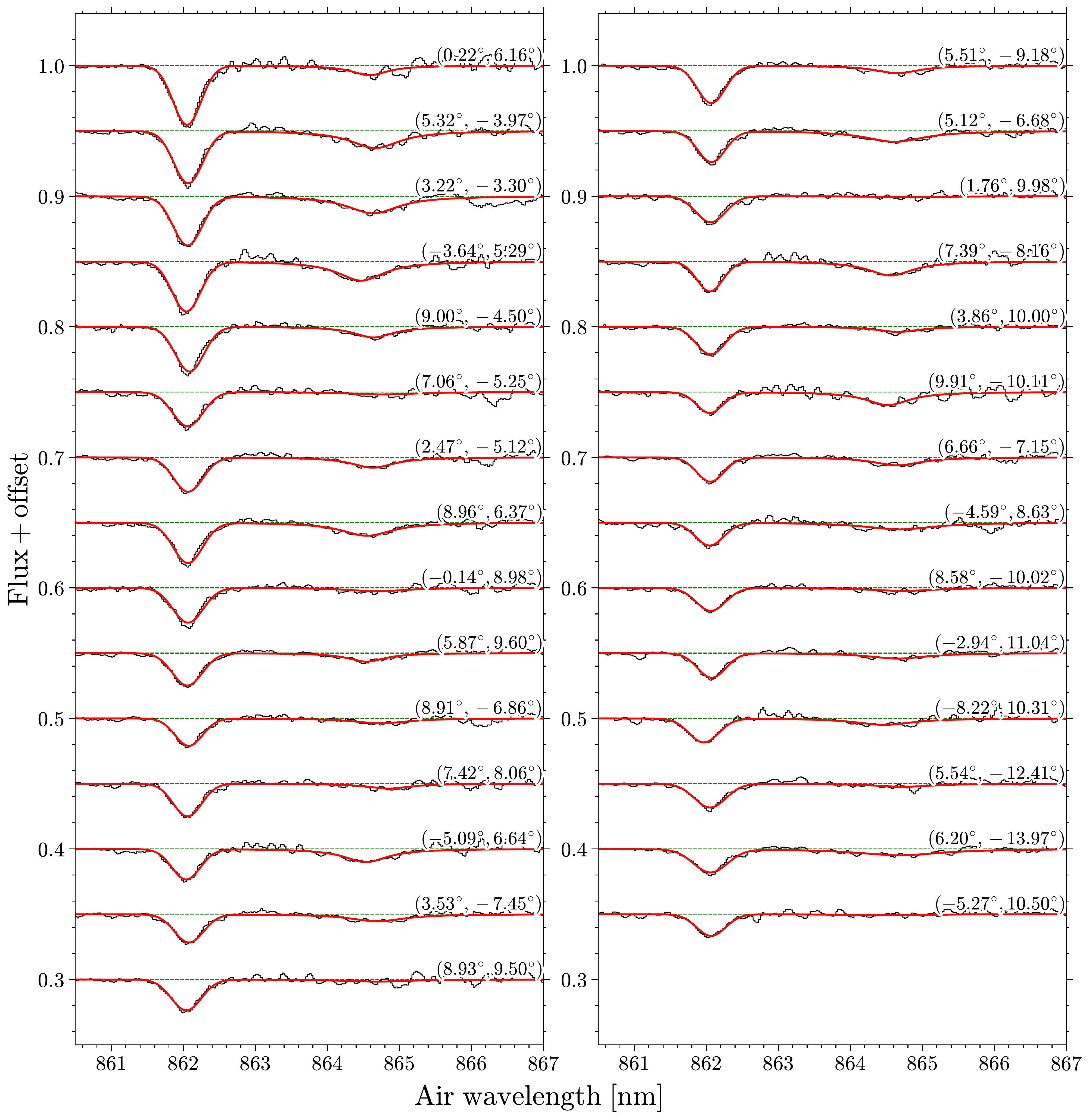}
    \caption{The same as Fig. \ref{fig:recover-4428}, but for DIB\,$\lambda$862.1.}
    \label{fig:recover-862.10}
\end{figure*}


\bsp	
\label{lastpage}

\clearpage

\end{CJK*}
\end{document}